\documentclass[preprint,12pt,authoryear]{elsarticle}

\usepackage{amssymb}
\usepackage{amsmath}
\usepackage{amsthm}
\usepackage{tabularx}
\usepackage[a4paper]{geometry}
\usepackage{lipsum}
\usepackage{multicol}
\usepackage{changepage}
\usepackage{multirow}
\usepackage{booktabs}
\usepackage{graphicx}
\usepackage{float}
\usepackage{xcolor}
\usepackage{lineno}
\usepackage{hyperref}
\usepackage{colortbl} 
\usepackage[normalem]{ulem} 
\usepackage{setspace}

\hypersetup{
    unicode=false,          
    pdftoolbar=true,        
    pdfmenubar=true,        
    pdffitwindow=false,     
    pdfstartview={FitH},    
    pdftitle={},    
    pdfauthor={},  
    pdfsubject={},   
    pdfcreator={},   
    pdfproducer={},  
    pdfnewwindow=true,     
    colorlinks=true,       
    linkcolor={blue},      
    citecolor=blue,        
    filecolor=blue,        
    urlcolor=blue          
}

\journal{}

\makeatletter
\def\ps@pprintTitle{%
 \let\@oddhead\@empty
 \let\@evenhead\@empty
 \let\@oddfoot\@empty
 \let\@evenfoot\@empty
}
\makeatother

\begin{document}

\begin{frontmatter}



\title{Fire severity and recovery across Europe: insights from forest diversity and landscape metrics}

\affiliation[inrae]{
    organization={INRAE, National Research Institute on Agriculture, Food \& the Environment, UMR TETIS}, 
    city={Montpellier},
    postcode={34090}, 
    country={France}
}

\affiliation[inria]{
    organization={INRIA, EVERGREEN, University of Montpellier},
    city={Montpellier},
    postcode={34090}, 
    country={France}
}

\author[inrae]{Eatidal Amin}
\author[inrae,inria]{C\'assio Fraga-Dantas}
\author[inrae,inria]{Dino Ienco}
\author[inrae]{Samuel Alleaume}
\author[inrae]{Sandra Luque}

\doublespacing

\begin{abstract}
In recent decades, European forests have faced an increased incidence of fire disturbances. This phenomenon is likely to persist, given the rising frequency of extreme events expected in the future. Estimating canopy recovery time after disturbance serves as a critical assessment for understanding forest resilience, which can ultimately help determine the ability of forests to regain their capacity to provide essential ecosystem services. This study estimated fire severity and post-disturbance recovery in European forests using a remote sensing–based time series approach. MODIS Leaf Area Index (LAI) time series data were used to track the evolution of vegetation cover over burned areas from 2001 to 2024. Fire severity was defined relative to pre-disturbance conditions by comparing vegetation status before and after fire events. Recovery intervals were determined from temporal evolution of vegetation greening as the duration required to reach the pre-disturbance LAI baseline. Furthermore, this study analyzed the severity and recovery indicators in relation to forest species diversity and landscape heterogeneity metrics across Europe, offering valuable insights into the spatial variability of forest response dynamics across diverse forest ecosystems across Europe. Results revealed a consistent pattern across vegetation cover types: higher forest species diversity and greater landscape shape complexity were associated with lower fire severity and, notably, shorter recovery times following fire disturbance.

\end{abstract}





\begin{keyword}
forest disturbances \sep remote sensing \sep landscape metrics \sep GWR \sep time series \sep forest fires \sep severity and post-fire recovery \sep LAI

\end{keyword}

\end{frontmatter}


\section{Introduction}
\doublespacing

European forests have faced increasing disturbances in recent decades, a trend expected to worsen with the growing frequency and intensity of extreme events \citep{sommerfeld2018,ellis2022}. This intensification of disturbance regimes across Europe is concerning due to its potential to disrupt forest management planning, the consistent and sustainable supply of timber, the carbon sequestration capacity, and the provision of multiple ecosystem services \citep{mayer2017,seidl2014,lecina2024}. Particularly, disturbance events significantly drive forest ecosystem dynamics by abruptly modifying their functioning and altering forest production resources, composition, water regulation, soil protection, and biodiversity conservation \citep{ceccherini2020, mouillot2013,driscoll2021}. Among the various threats to forest ecosystems, fire stands as the second most significant disturbance in European forests \citep{bonannella2024}, accounting for 24\% of the total timber volume damaged between 1950 and 2019 \citep{patacca2023}.

To effectively address the growing prevalence of fire risk marked by an increasing frequency and intensity \citep{gonzalez2016,jones2022}, robust data on the spatial and temporal distribution of fire occurrences is essential \citep{ganteaume2013}. Such data enables the assessment of fire severity and forest recovery, which can further contribute to a clearer understanding of the environmental factors influencing these disturbance events. This knowledge, in turn, facilitates the development of adaptive mitigation management measures, enhancing long-term forest resilience, minimizing tree mortality, and ensuring sustainable harvesting. On one hand, severity describes the magnitude of the loss or disruption that a disturbance causes within the ecosystem structure and function \citep{white2017}. On the other hand, estimating post-disturbance recovery time, as the period required for an ecosystem to return to its pre-disturbance state, involving vegetation properties and functioning conditions, offers valuable insights into forest resilience \citep{ingrisch2018,senf2022}.

Advances in cloud computing platforms and long-term Earth Observation satellite data have enabled large-scale mapping of post-disturbance dynamics, from regional to continental extents \citep{senf2021,forzieri2021}. Satellite-based monitoring has has been extensively employed for assessing disturbance severity and recovery by utilizing spectral indices or biophysical variables to track the temporal evolution of canopy vegetation \citep{pickell2016,garcia2019,nole2022}. Commonly, the former primarily involves calculating the difference or magnitude range between pre- and post-disturbance spectral indices data, while the latter mainly relies on defining recovery thresholds \citep{hermosilla2019,frazier2018,mandl2024}.

While substantial advances in studying disturbance severity and recovery related to indicators of climatic drivers \citep{forzieri2024, ermitao2024}, forest structure \citep{cerioni2024}, and biogeographical variability \citep{nole2022} have been recently made, comprehension of the overarching patterns and the landscape factors involved in the forest recovery processes across large spatial extents, such as the whole of Europe, invites to further investigation. This knowledge gap extends to how forests recover over space, in relation to different species compositions and ecosystem spatial patterns \citep{moreira2011}. In particular, landscape heterogeneity, stemming from variations in vegetation composition and structure, has been theorized as a key factor in enhancing forest resistance to natural disturbances \citep{jactel2017,jactel2021,meigs2018}. However, large-scale empirical studies exploring the relationship between landscape heterogeneity, forest species diversity and disturbance severity, as well as post-disturbance recovery, are still lacking. Therefore, understanding these dynamics across Europe would be highly beneficial for informing EU forest adaptation strategies and guiding regional and national investment priorities towards proposing relevant actions for improving forest resilience.

Forest fires have the most extensive and spatially detailed database in Europe, spanning the past two decades. Using satellite remote sensing time series, this study introduces indicators to quantify fire severity and recovery across various forest types in Europe. Recovery is assessed via a threshold-based method that determines the time needed for vegetation to return to pre-fire conditions, ensuring adaptability across diverse ecological contexts. The study aims to identify continental-scale patterns in severity and recovery, revealing how responses vary with vegetation cover. Ultimately, these patterns support the overarching goal of understanding the relationship between forest species diversity, landscape heterogeneity, and fire disturbance, as characterized by the derived severity and recovery metrics.

\section{Materials and methods}

\subsection{Disturbance and remote sensing data}

This study focuses on fire as a major forest disturbance across Europe, using spatially explicit vector data on burned areas from 2001 to 2024 to assess its impact on diverse forest landscapes (Table \ref{tab:DRSdata}). Fire occurrence data were sourced from the European Forest Fire Information System (EFFIS, \url{https://forest-fire.emergency.copernicus.eu/}). To analyze the temporal dynamics of vegetation across Europe, a time series of 8-day composite MODIS Leaf Area Index (MODIS LAI) images, covering the period from the beginning of 2001 to the end of 2024, was used. LAI served as a descriptor of the vegetation condition, reflecting the seasonal phenological evolution of photosynthetically active vegetation \citep{yan2019}. The CORINE Land Cover (CLC) level 3 dataset (\url{https://land.copernicus.eu/}), reclassified into 13 distinct landscape classes (Table \ref{tab:LCReclass}), served two primary functions in this study: (i) it provided the underlying land cover necessary to process and link recovery and severity indicators specifically for each forest ecosystem; and (ii) it enabled the estimation of landscape metrics across Europe. Finally, species diversity throughout Europe was quantified using forest species data from the European Atlas of Forest Tree Species (JRC Tree Atlas) \citep{TreeAtlas}.

\begin{table}[H]
\begin{center}
\caption{Remote sensing based products used in the time series processing framework proposed for analyzing fire disturbance dynamics at European level.} 
\resizebox{0.75\textwidth}{!}{
\begin{tabular}{lcc}
\hline
Data product & Spatial resolution &  Temporal coverage \\ 
\hline
MODIS LAI & 500 m & 2001-2024 \\ 
CORINE Land Cover & 100 m & 2018 \\
JRC Tree Atlas & 1 km & 2016 \\
\hline
\end{tabular}
}
\label{tab:DRSdata}
\end{center}
\end{table}

\subsection{Processing framework: severity and recovery of disturbance}

This study developed a remote sensing framework to derive fire disturbance severity and vegetation recovery. Fire severity was determined by comparing pre- and post-disturbance greenness status, while recovery was defined as the time for canopy cover to return to its pre-disturbance baseline. Burned area data from the EFFIS database were used to identify the location and timing of the fire event, with all analyses conducted per pixel within the affected area.

For each pixel within a disturbed area, a 1-year window Moving Average $MA(t)_{Baseline}$ filter was applied to the MODIS LAI time series, $LAI (t)$, considering all available years before the fire date $(T_{fire})$:

\begin{equation}
MA (t)_{Baseline} = \frac{1}{365} \sum_{i=t-364}^{t} LAI (i), 
\text{with } t \in T_{b} = \{T_{0},..., T_{fire}\}
\end{equation}

Using the same temporal window, the moving average $(MA)$ time series processing was performed forward, from $T_{fire}$ to the end of the time series, $T_{N}$:

\begin{equation}
MA (t) = \frac{1}{365} \sum_{i=t}^{t+364} LAI (i), 
\text{with } t \in T_{p} = \{T_{fire},..., T_{N}\}
\end{equation}

The recovery date threshold was established as the pre-disturbance mean $\overline{MA}_{Baseline}$, minus one standard deviation, $\text{std}(MA_{Baseline})$:
\begin{equation}
\text{Threshold} = \overline{MA}_{Baseline} - \text{std}(MA_{Baseline})
\end{equation}
Recovery following a disturbance was then considered achieved when $MA(t)$ reached and remained at or above this threshold for a continuous period of one year:
\begin{equation}
\begin{alignedat}{2}
\text{Recovery date } T^* =& \min\!\left(t \in \{T_{fire}, ..., T_N\} \right) \text{ such that }\\
 MA(t') \geq& \text{ Threshold} ~~\forall\, t' \in \{t, ..., t+364\} 
\end{alignedat}
\end{equation}

The recovery date was then identified as the first day of the time period meeting such a condition. The fire impact was assessed by calculating the disturbance severity, as the relative difference between the LAI value after and before the disturbance, thus representing the percentage loss of vegetation due to the disturbance event. 

All initial calculations and image processing were performed at the MODIS original spatial resolution in Google Earth Engine \citep{gorelick2017}. Resulting severity and recovery maps were then exported and further processed at 1km spatial resolution.

\subsection{Forest species diversity and landscape metrics}

Forest tree species diversity was calculated using the European Atlas of Forest Tree Species \citep{TreeAtlas}. This dataset includes a relative probability of presence map for each species, offering spatially explicit insights into their potential distribution. Specifically, the Shannon diversity index \citep{maurer2011} was computed at the pixel level, leveraging the relative probability of presence maps of all available species. These maps were summed, normalized by the total sum to represent relative abundances, and finally used as input to estimate forest species diversity. 

Spatial patterns were quantified and analyzed using various landscape shape and complexity metrics from the PyLandStats library \citep{bosch2019}. These metrics were calculated from the reclassified 2018 CORINE Land Cover map, comprising 13 final classes (Table \ref{tab:LCReclass}). A 5 $\times$ 5 km focal moving window centered on each processed pixel was applied to capture the spatial complexity and geometric attributes of surrounding land cover types. The resulting map was then spatially resampled to a 1 km spatial resolution. The PyLandStats metrics are primarily computed based on their definitions in FRAGSTATS \citep{mcgarigal2023} (see Section \ref{sec:lmetrics_formulas}). The specific metrics used in this study were:

\begin{itemize}

    \item[-] Landscape shape index (LSI). Indicates patch shape complexity, with higher values reflecting more irregular and fragmented landscapes.
    
    \item[-] Effective mesh size (MESH). Measures landscape connectivity. Low values indicate higher fragmentation, while high values suggest large, cohesive patches. 

    \item[-] Largest patch index (LPI). Represents dominance of the largest patch. Low values imply balanced distribution, while high values denote patch dominance.    
    
\end{itemize}

\subsection{Species diversity and landscape in spatial assessment of fire severity and recovery}

To spatially evaluate the relationship between fire disturbance, species diversity and landscape complexity, Geographically Weighted Regression (GWR), a statistical technique designed to capture the variation of processes across space \citep{fotheringham2009}, was used. GWR enables localized modeling, improving model fit and reducing spatial autocorrelation compared to global regression models that assume spatially uniform effects. GWR has been widely used to identify geographic variation and disentangle multiple ecological drivers, including vegetation dynamics, land-use impacts, and fire regimes \citep{xue2023, nunes2016, harris2010}.The analysis in this study was conducted using the mgwr Python library \citep{oshan2019}.

GWR essentially calibrates separate local linear regression models at each location using data from neighboring pixels, where observations are weighted by their distance:

\begin{equation}
y_i = \beta_{i0} + \sum_{k=1}^{p} \beta_{ik} x_{ik} + \epsilon_i,\quad i = 1, \ldots, n
\end{equation}

where the index $i$ refers to the spatial coordinates $(u_i, v_i)$, indicating the geographic position of each regression point, $y_i$ denotes the dependent variable at location $i$, $\beta_{i0}$ is the location-specific intercept, $x_{ik}$ represents the $k$-th explanatory variable at location $i$, $\beta_{ik}$ is the corresponding regression coefficient for the $k$-th explanatory variable at location $i$, and $\epsilon_i$ is the random error term. Bandwidth calibration in GWR is crucial for regression sensitivity. Fixed kernels apply uniform distances, while adaptive kernels adjust to maintain a consistent neighbor count, better suited for clustered data like burned areas. This study employed an adaptive bi-square kernel, assigning zero weight beyond the bandwidth and optimizing it via AICc to improve local accuracy limiting the influence of distant points. The resulting spatial weighting supports locally adaptive models and the mapping of regression parameters.

\section{Results}
\doublespacing

\subsection{Mapping disturbance severity and recovery across Europe}

\begin{figure}[H]
    \centering
    \begin{minipage}{0.45\textwidth}
        \centering
        \includegraphics[width=\textwidth]{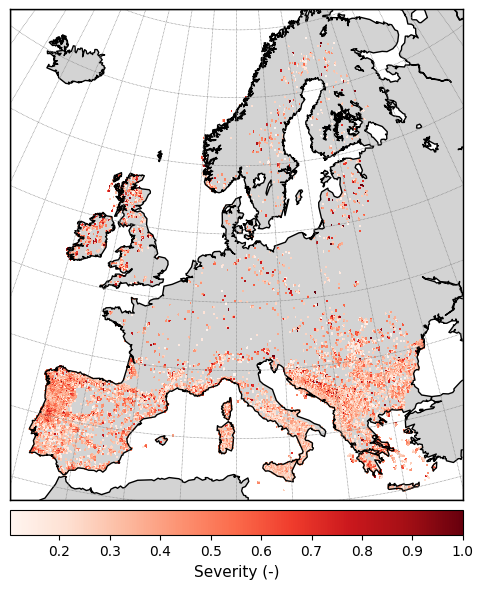} 
    \end{minipage}
    \begin{minipage}{0.27\textwidth}
        \centering
        \includegraphics[width=\textwidth]{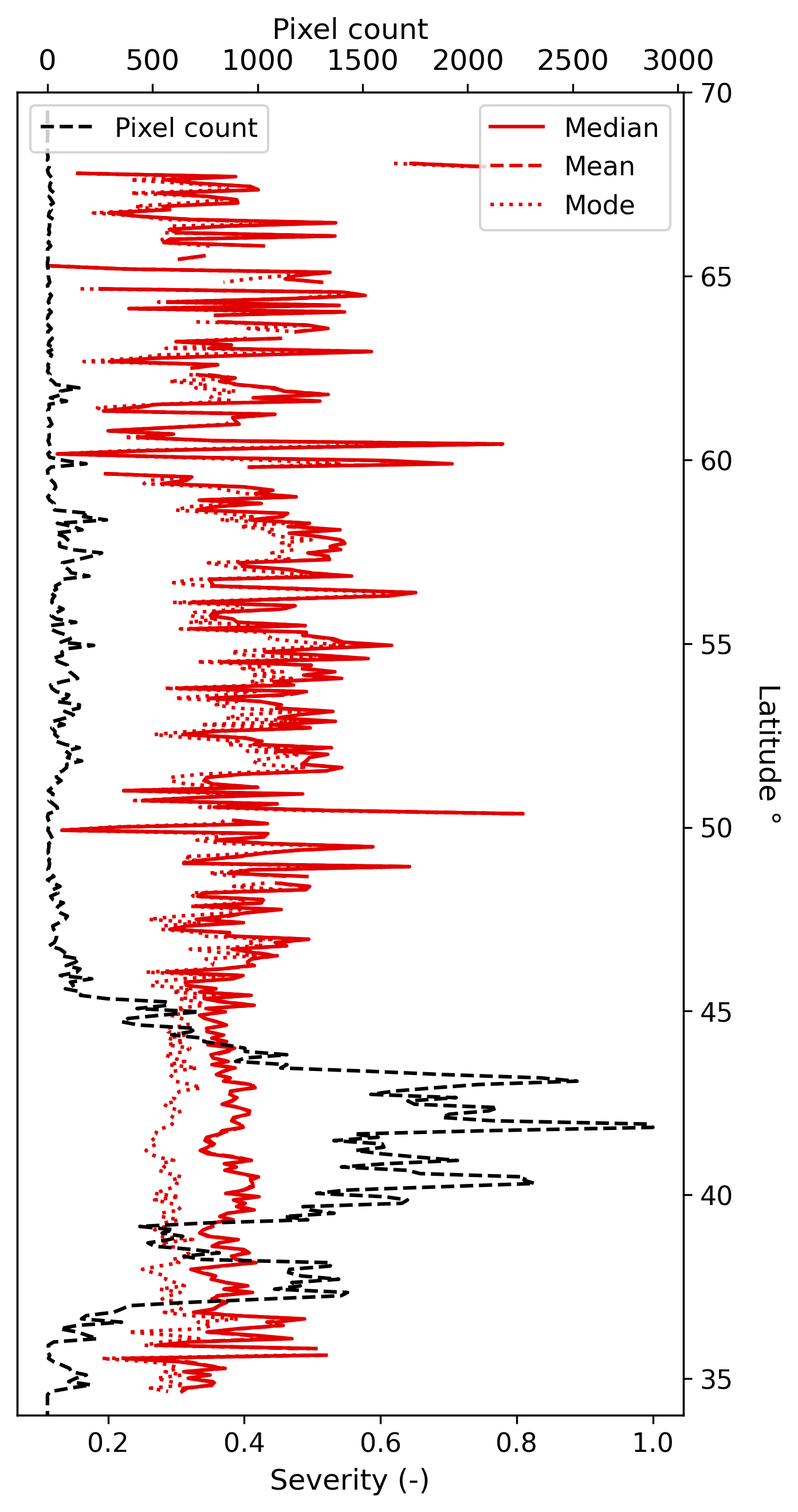} 
        \vspace{0.47cm}
    \end{minipage}

    \vspace{0.05cm}
    \begin{minipage}{0.32\textwidth}
        \centering
        \includegraphics[width=\textwidth]{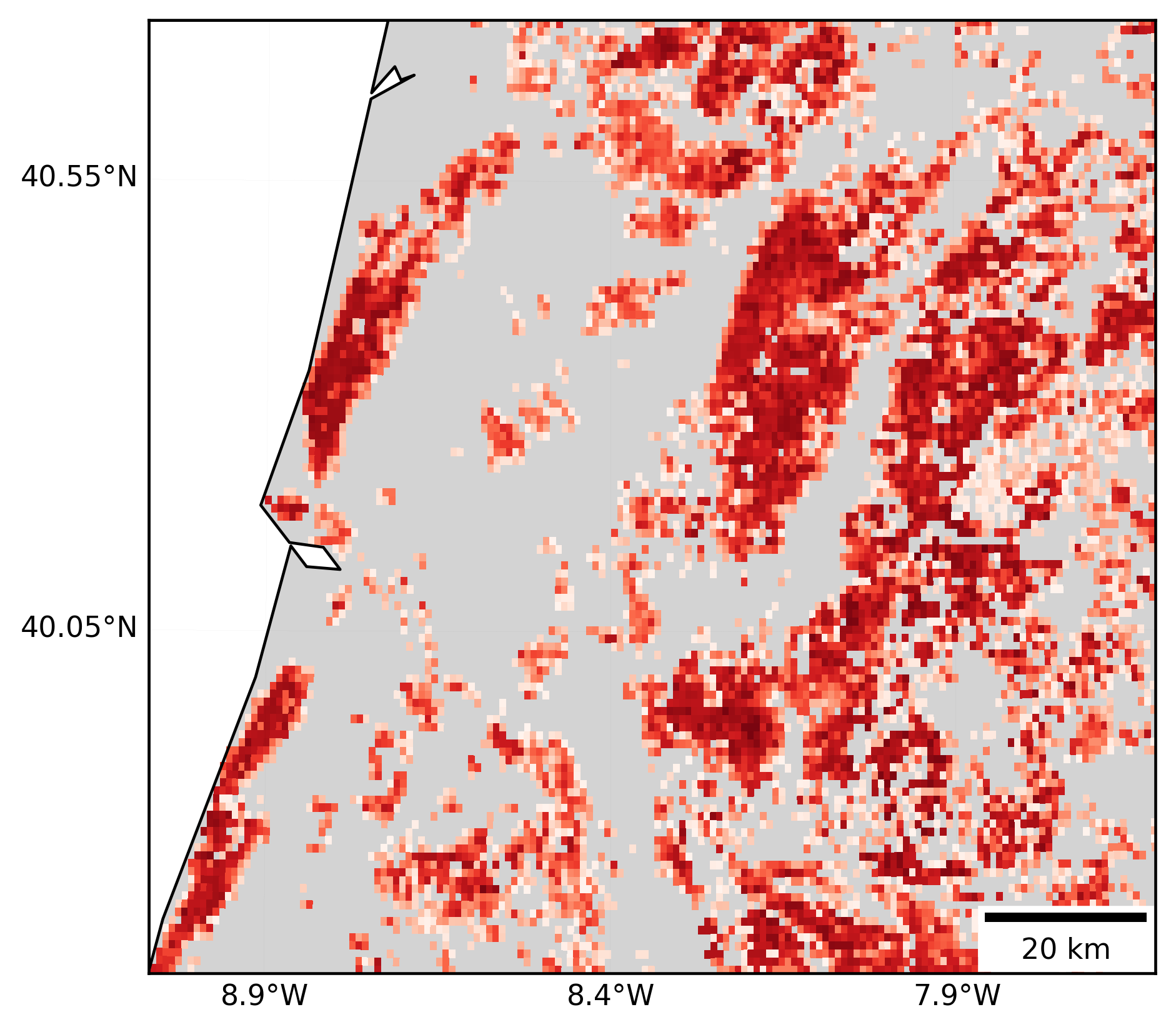}
    \end{minipage}
    \hfill
    \begin{minipage}{0.32\textwidth}
        \centering
        \scalebox{1.15}{\includegraphics[width=\textwidth]{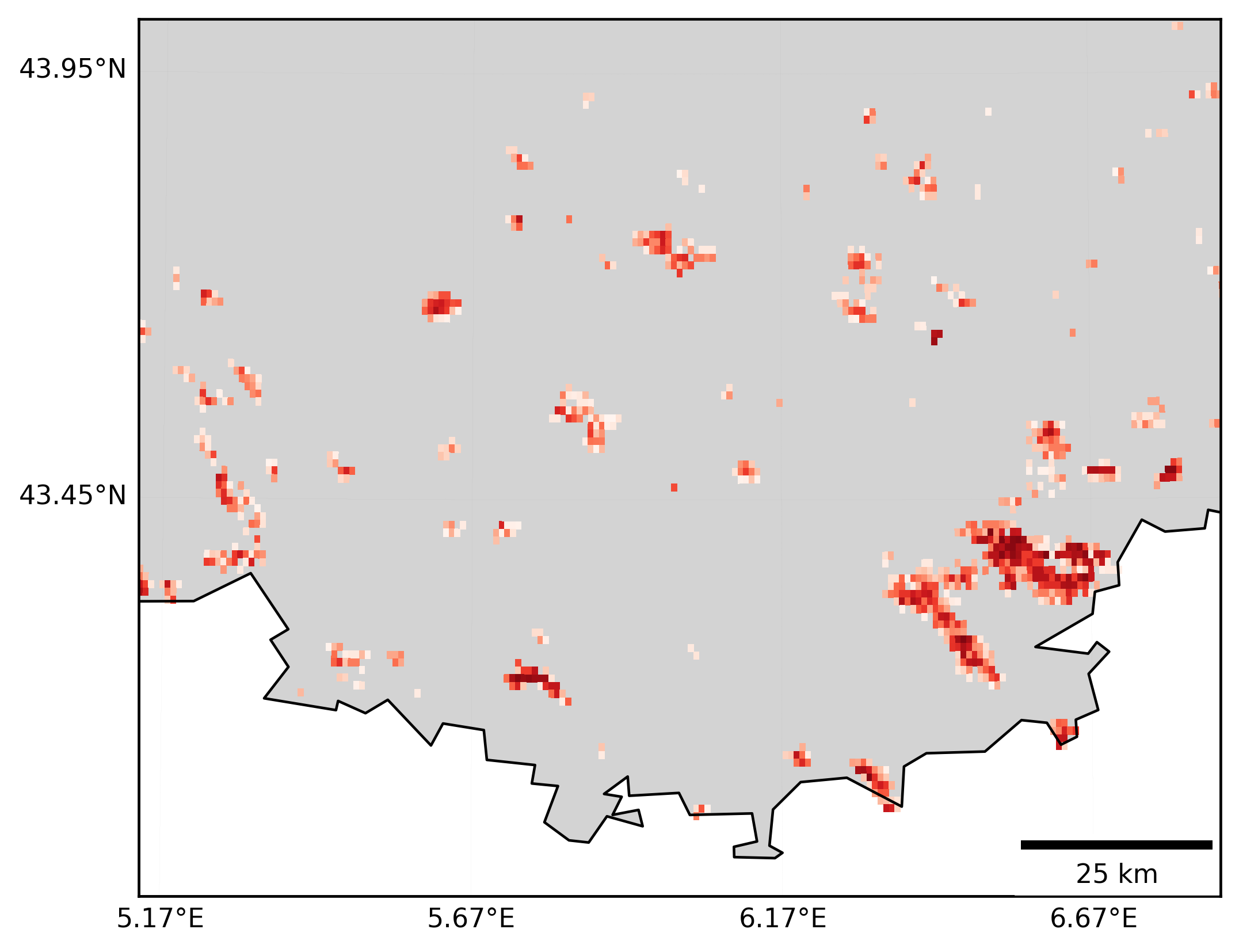}}
    \end{minipage}
    \hfill
    \begin{minipage}{0.32\textwidth}
        \centering
        \scalebox{0.745}{\includegraphics[width=\textwidth]{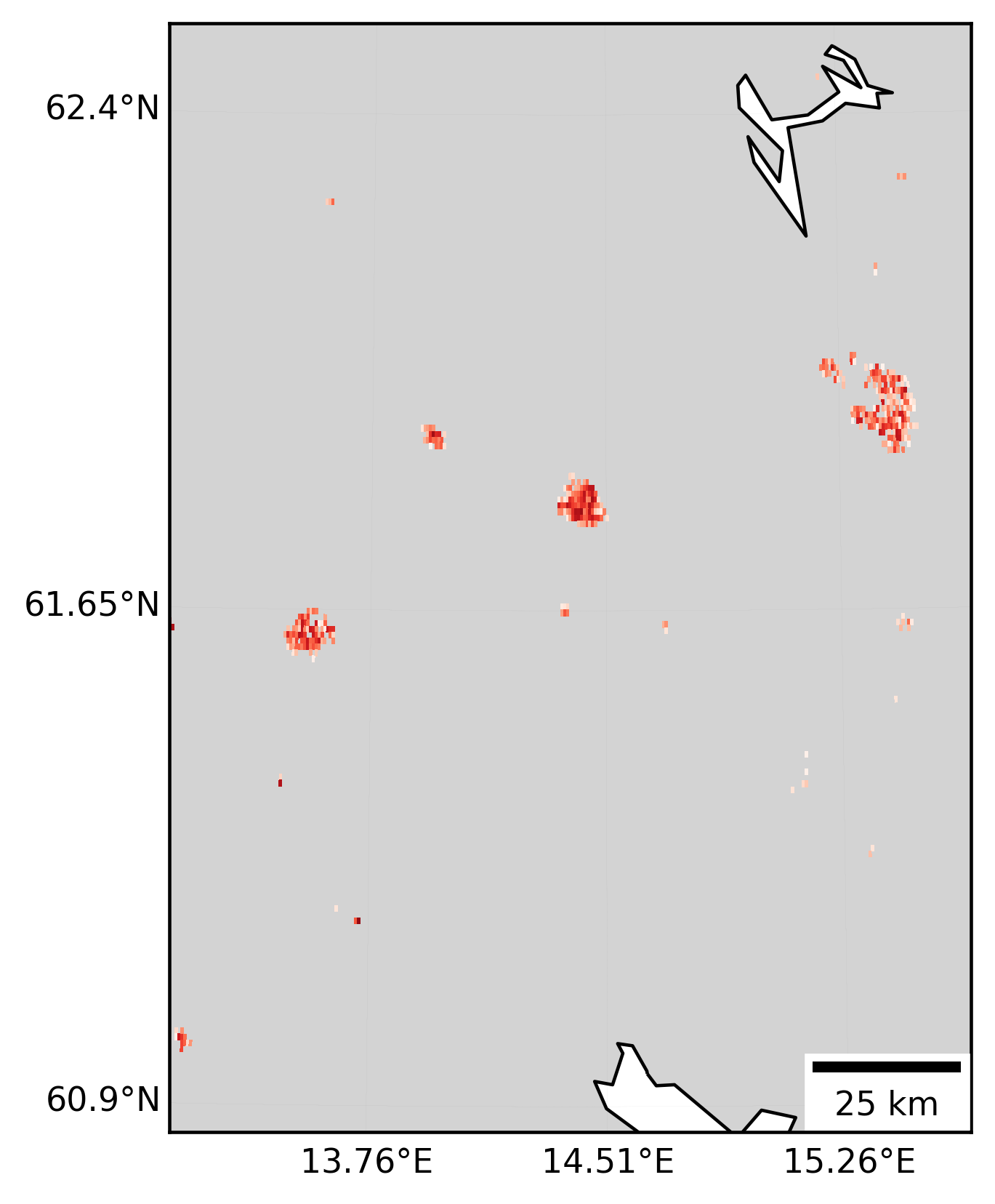}}
    \end{minipage}

    \caption{Fire severity across Europe from 2001 to 2024 fires. Map values are averaged over a 10$\times$10 km moving window to enhance visual interpretability. The vertical histogram displays pixel counts corresponding to burned areas, along with the mean, median and mode severity values distributed along the latitudinal gradient.  The maps below display zoomed-in views of representative regions in Portugal, France, and Sweden (from left to right).} 
    \label{fig:Sevmap}
\end{figure}

\begin{figure}[H]
    \centering
    \begin{minipage}{0.45\textwidth}
        \centering
        \includegraphics[width=\textwidth]{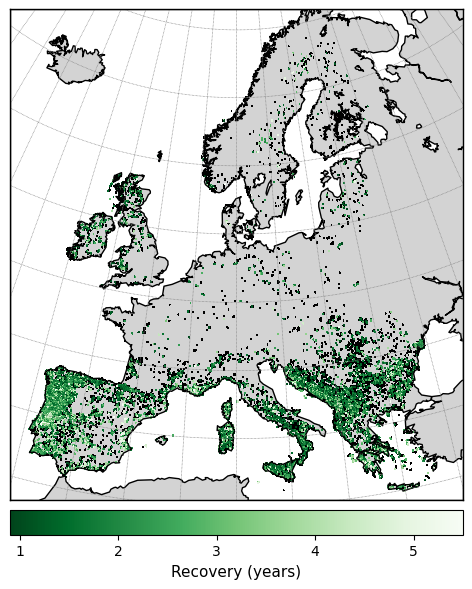} 
    \end{minipage}
    \begin{minipage}{0.27\textwidth}
        \centering
        \includegraphics[width=\textwidth]{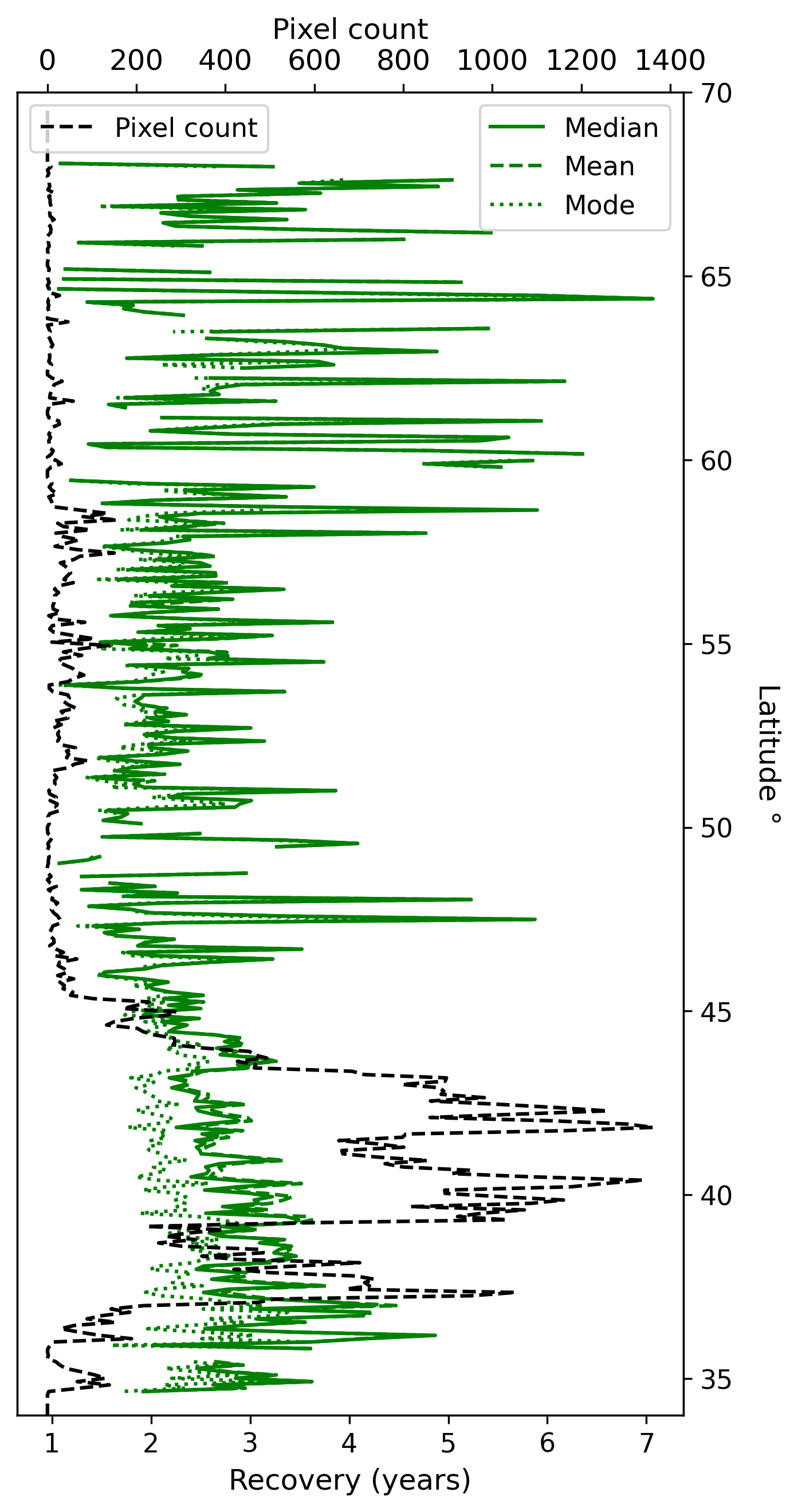} 
        \vspace{0.47cm}
    \end{minipage}    

    \vspace{0.05cm}
    \begin{minipage}{0.32\textwidth}
        \centering
        \includegraphics[width=\textwidth]{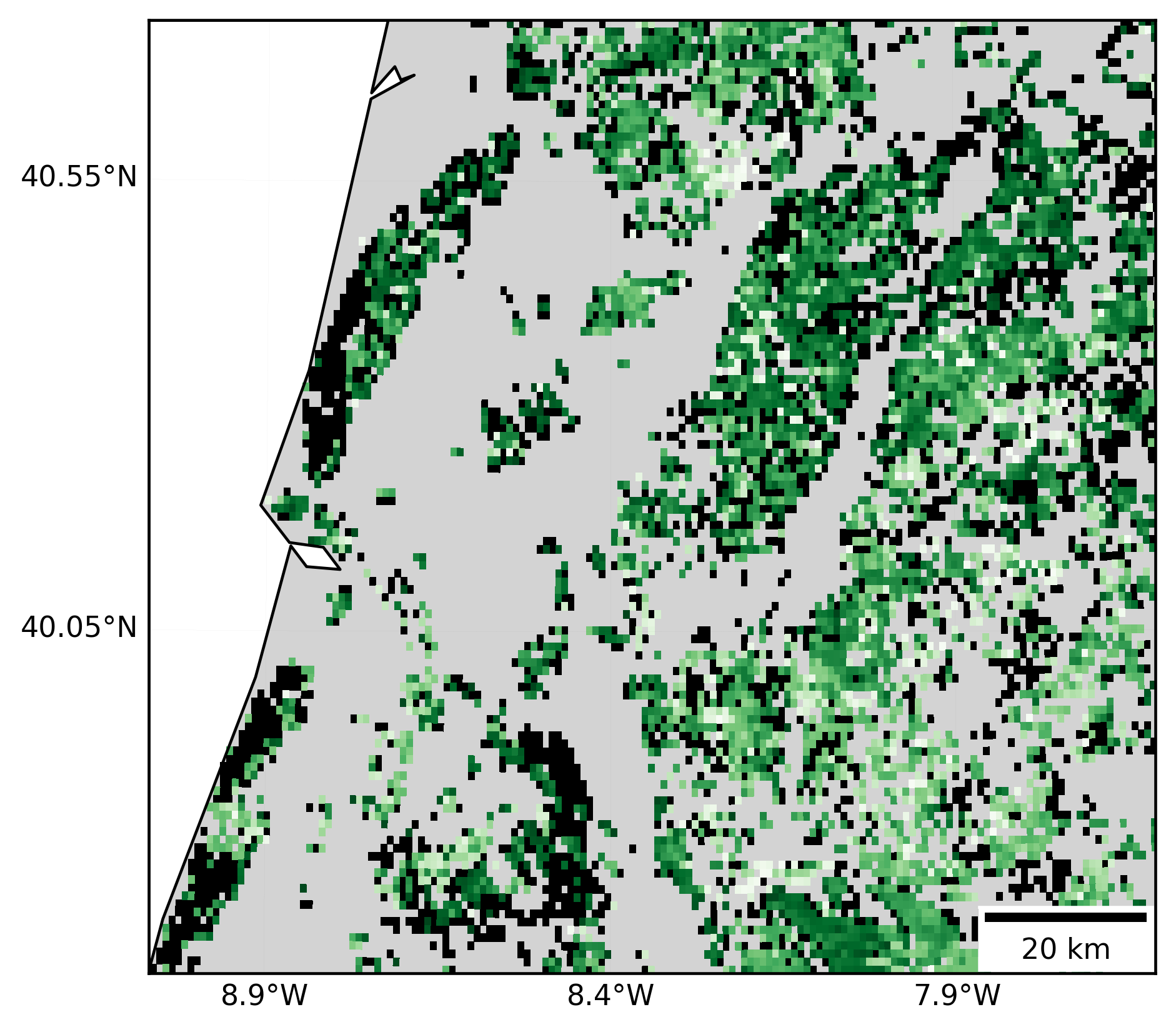}
    \end{minipage}
    \hfill
    \begin{minipage}{0.32\textwidth}
        \centering
        \scalebox{1.15}{\includegraphics[width=\textwidth]{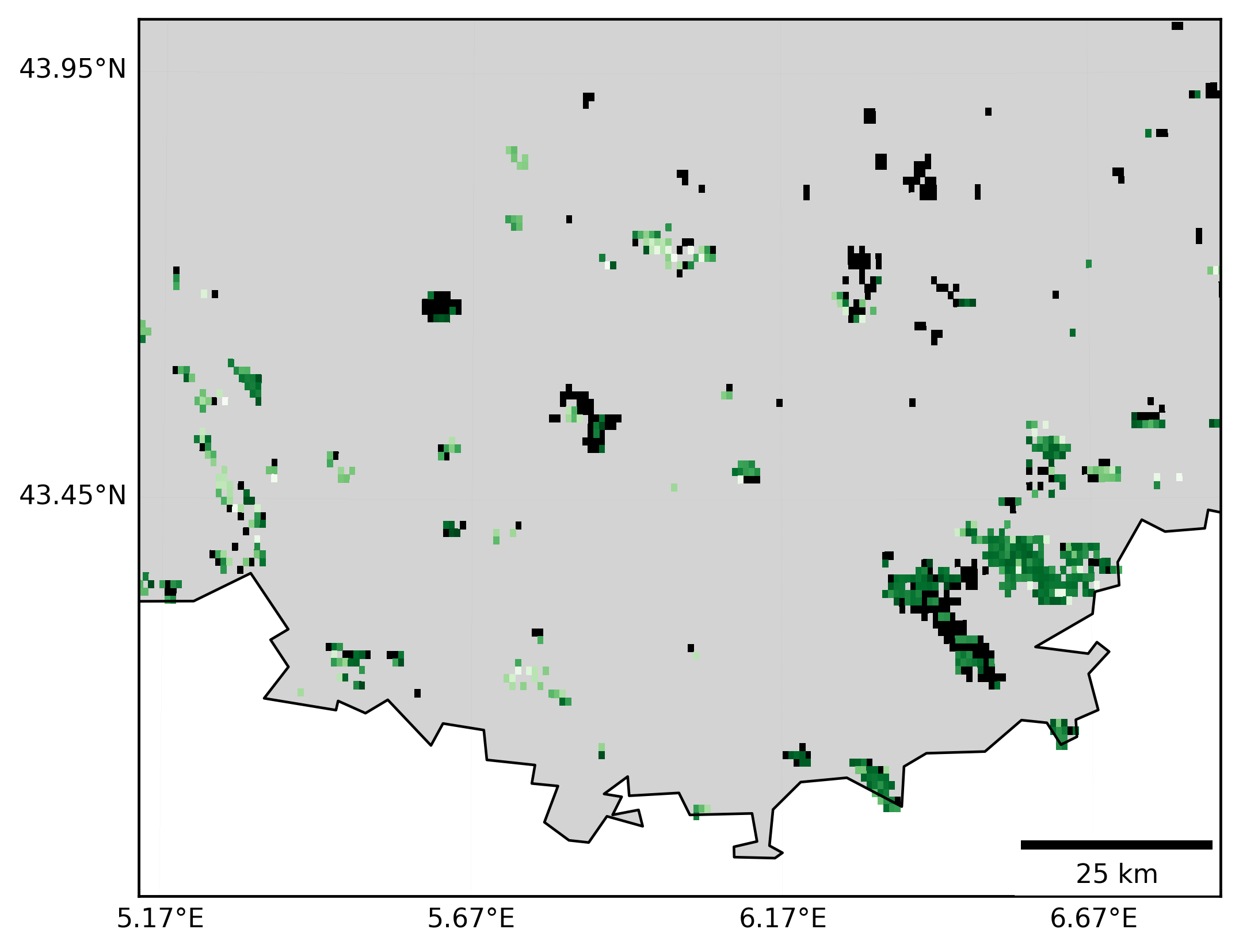}}
    \end{minipage}
    \hfill
    \begin{minipage}{0.32\textwidth}
        \centering
        \scalebox{0.745}{\includegraphics[width=\textwidth]{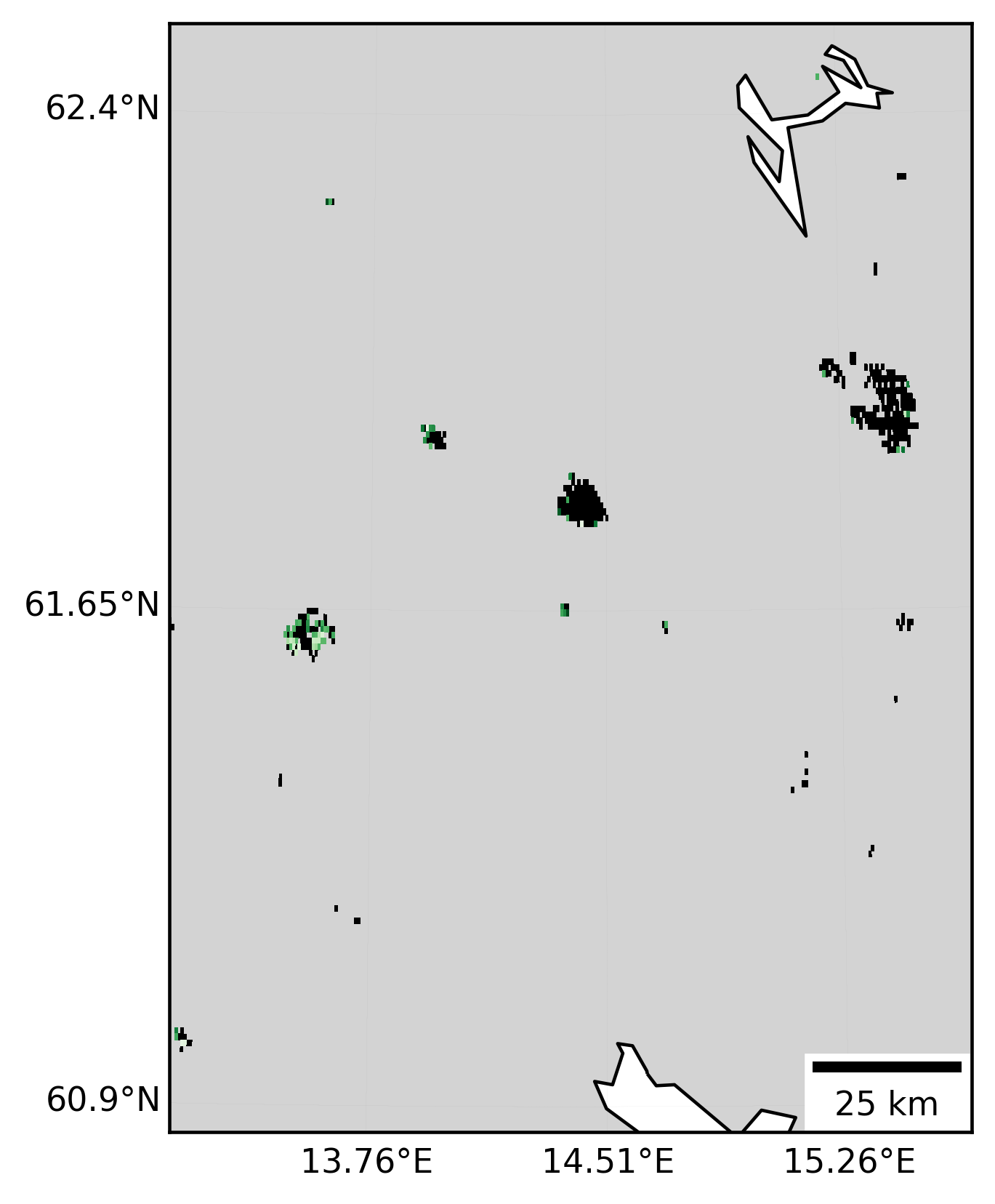}}
    \end{minipage}
    
    \caption{Fire Recovery time across Europe from 2001 to 2024 fires. Black pixels represent unrecovered pixels. Map values are averaged over a 10$\times$10 km moving window to enhance visual interpretability. The vertical histogram displays pixel counts corresponding to burned areas, along with the mean, median and mode recovery values distributed along the latitudinal gradient. The maps below display zoomed-in views of representative regions in Portugal, France, and Sweden (from left to right).} 
    \label{fig:Recmap}
\end{figure}

The image-based processing framework enabled mapping the severity of fire disturbance and subsequent recovery across Europe for fires from 2001 to 2024 (Figures \ref{fig:Sevmap}, \ref{fig:Recmap}). Fire incidence was highest in Southern European countries, with particularly severe fires mostly occurring in Portugal, Spain and Greece. While lower latitudes experience significantly more fires, higher latitudes, though fewer in number, exhibit a higher proportion of severe fires. Overall, lower and mid latitudes typically experience faster recovery periods than the slower recoveries seen in higher latitudes, including boreal regions. Notably, the unrecovered pixels represent 36.22\%, with most corresponding to recent fires.

\subsection{Fire disturbance by land cover}

\begin{figure}[H]
    \centering
    \begin{minipage}{0.5\textwidth}
        \centering
        \includegraphics[width=\textwidth, trim=30 0 0 0, clip]{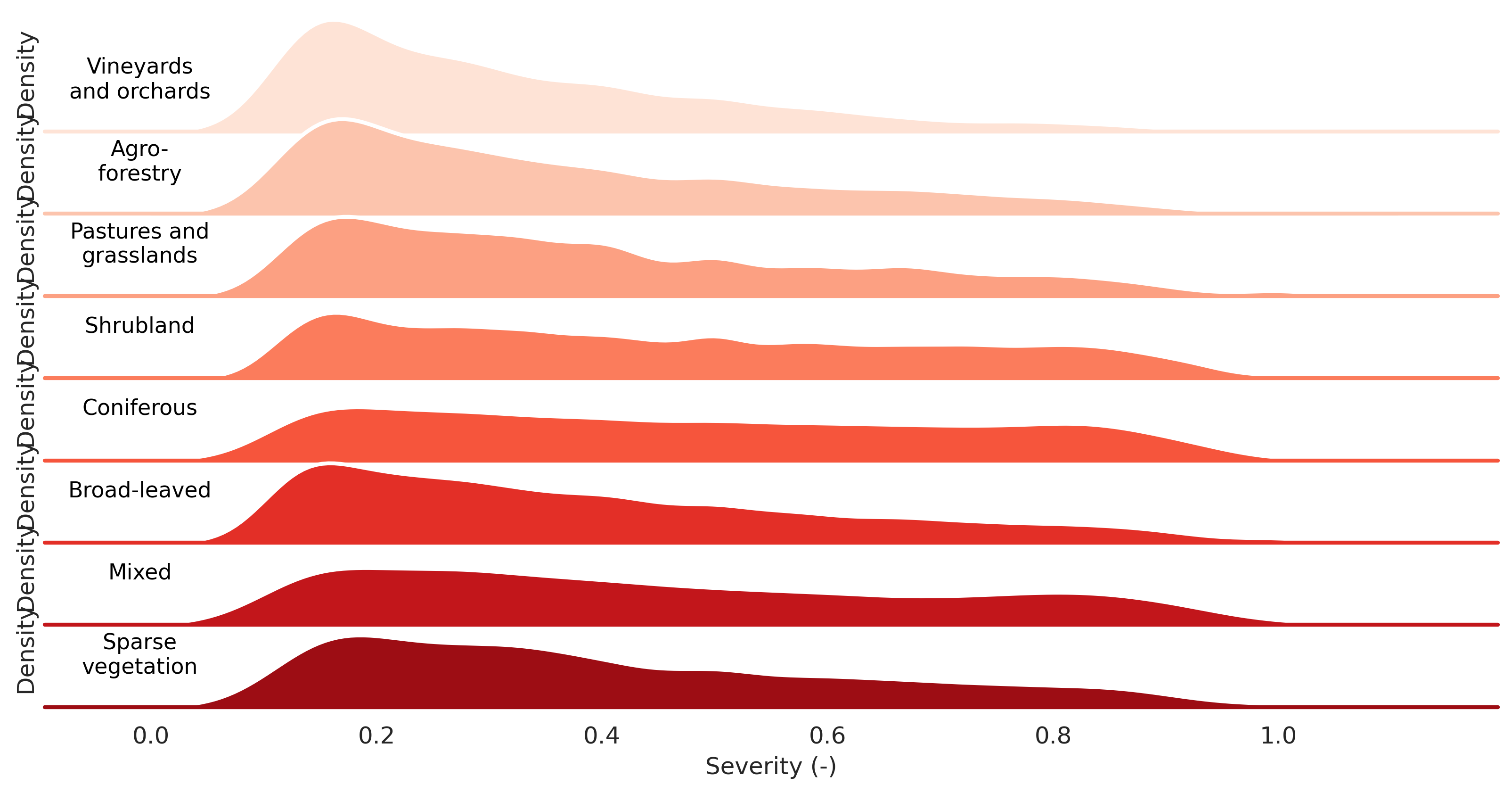} 
    \end{minipage}
    \begin{minipage}{0.5\textwidth}
        \centering
        \includegraphics[width=\textwidth, trim=18 0 0 0, clip]{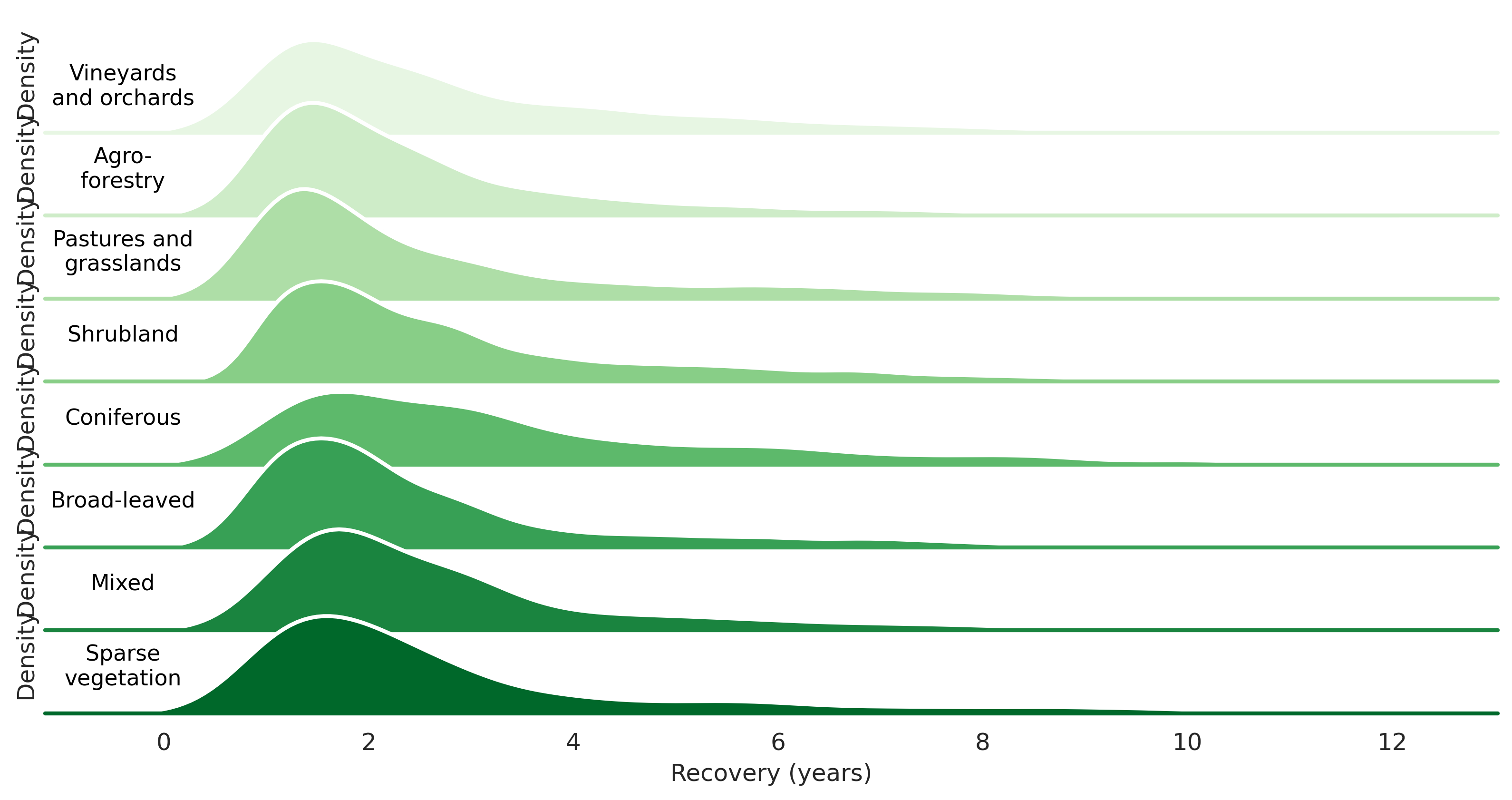}        
    \end{minipage}
    \caption{Distribution of fire severity (left) and recovery (right) per land cover class over Europe from 2001 to 2024 fires.}
    \label{fig:sev-rec-distrib}
\end{figure}

Recovery intervals and disturbance severity metrics were analyzed across a wide range of European forested landscapes to assess the influence of spatial heterogeneity and species composition on severity and post-fire recovery dynamics (Table \ref{tab:Disturbmetrics}). To avoid distortions in further mapping and processing, the lowest severity and recovery pixels were discarded using a minimum threshold (0.1 for severity and 0.9 years for recovery). These low values, likely caused by rasterization and pixel size limitations, did not reflect a typical disturbance trajectory in the time series. Analysis of fire incidence by land cover type (Figure \ref{fig:sev-rec-distrib}) showed that the most severe fires were more frequent in coniferous forests, closely followed by shrublands and mixed forests. Vineyards and agroforestry zones display comparable fire patterns, generally experiencing lower fire severity. Fire severity is also consistently low in pastures and broad-leaved forests. Mixed forests exhibited lower fire severity compared to coniferous forests and shorter recovery times than broad-leaved forests. Average recovery periods were longest in coniferous forests and shrublands. Similarly, higher minimum recovery periods were observed in coniferous and shrublands. Conversely, faster recovery times were observed in broad-leaved forest and agroforestry ecosystems. 


\begin{table}[H]
\begin{center}
\caption{Summary statistics of fire disturbance severity and recovery by land cover class: mean, median, standard deviation, quartile 25\%, quartile 75\%.} 
\resizebox{\textwidth}{!}{
\begin{tabular}{lccccc|ccccc}
\hline
\multirow{2}{*}{Land cover class} & \multicolumn{5}{c}{Severity (-)} & \multicolumn{5}{c}{Recovery (years)}\\
 & mean & median & std & q25 & q75 & mean & median & std & q25 & q75\\ 
\hline

Vineyards and orchards & 0.32 & 0.27 & 0.18 & 0.17 & 0.43
& 3.02 & 2.32 & 2.25  & 1.41  & 3.89\\

Agroforestry & 0.37 & 0.31 & 0.21 & 0.20 & 0.50 
& 2.77 & 2.00 &	2.27  &	1.39 &	3.21\\

Pastures and grasslands & 0.40 & 0.33 & 0.22 & 0.21 & 0.57
& 3.10	& 1.98	& 2.76	& 1.33	& 3.76\\

Shrubland & 0.46 & 0.43 & 0.24 & 0.25 & 0.67
& 3.36	& 2.39	& 2.77	& 1.58	& 4.09\\

Coniferous & 0.48  & 0.45  & 0.24  & 0.26 &	0.69
& 3.70	& 2.86	& 2.71	& 1.77	& 4.75\\

Broad-leaved & 0.40	& 0.33	& 0.23	& 0.20	& 0.55
& 2.91 & 2.01 & 2.50 &	1.42 & 3.18\\

Mixed & 0.46 & 0.41	& 0.25	& 0.25	& 0.66
& 3.00	& 2.25	& 2.26	& 1.63	& 3.37\\

Sparse vegetation & 0.41 & 0.36	& 0.22 & 0.22 & 0.57
& 3.14	& 2.10	& 2.75	& 1.45	& 3.57\\

\hline
\end{tabular}
}
\label{tab:Disturbmetrics}
\end{center}
\end{table}

\subsection{Fire disturbance dynamics: Forest species diversity and landscape metrics}

\begin{figure}[H]
    \centering
    \begin{minipage}{0.45\textwidth}
        \centering
        \includegraphics[width=\textwidth]{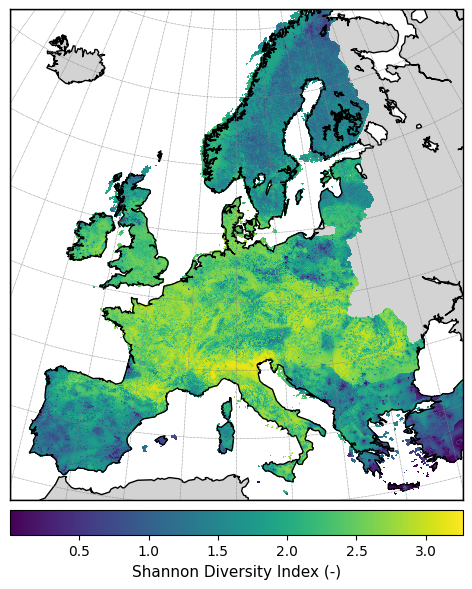} 
    \end{minipage}
    \begin{minipage}{0.27\textwidth}
        \centering
        \includegraphics[width=\textwidth]{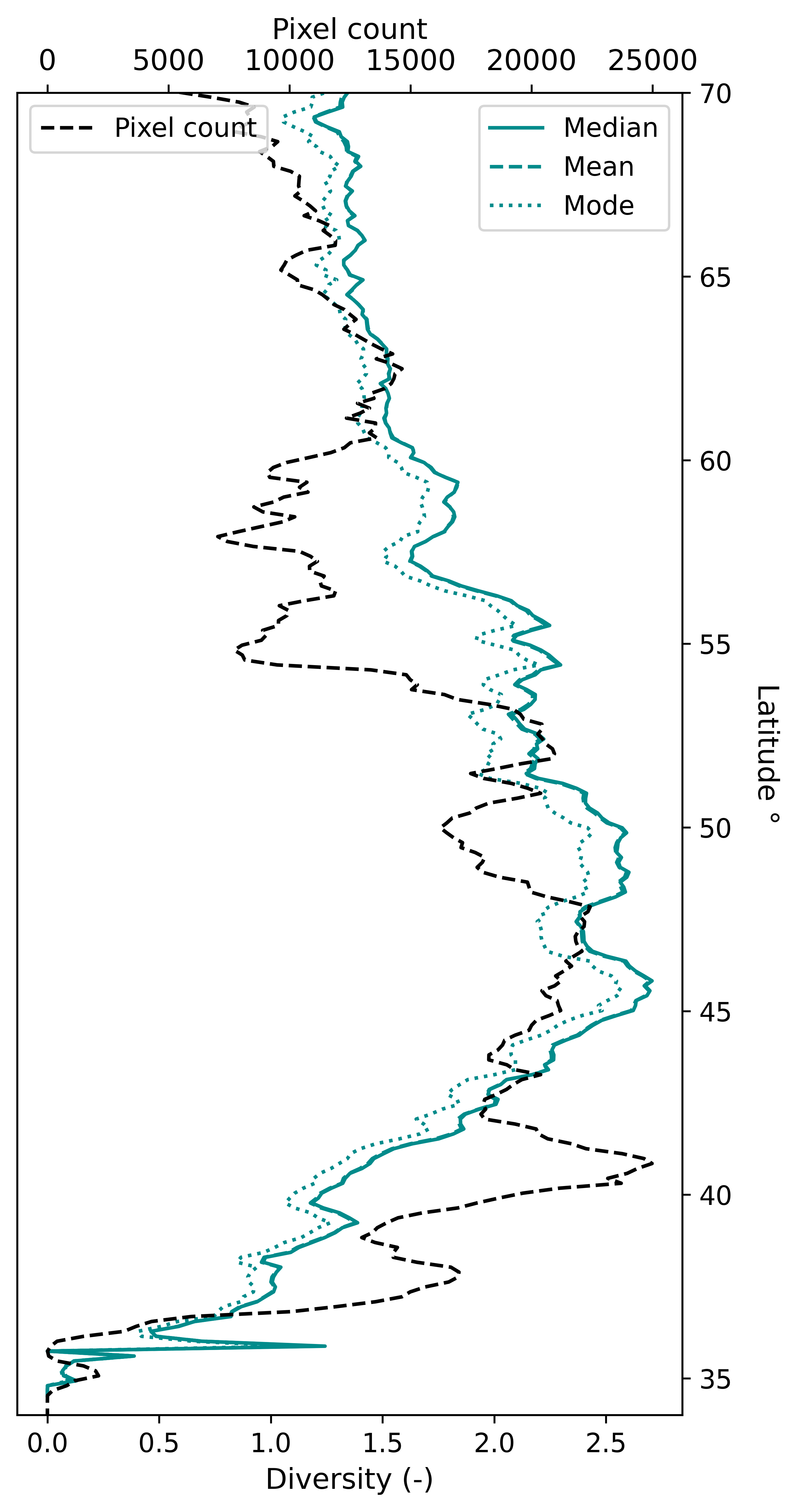} 
        \vspace{0.47cm}
    \end{minipage}
    \caption{Forest species diversity map corresponding to the Shannon diversity index calculated from JRC Tree Atlas data. The vertical histogram displays pixel counts corresponding to burned areas, along with the mean and median diversity values distributed along the latitudinal gradient.} 
    \label{fig:Divmap}
\end{figure}

\begin{figure}[H]
    \centering
    \begin{minipage}[c]{0.31\textwidth} 
        \centering
        LSI \
        \includegraphics[width=\textwidth,trim=0 20 0 0, clip]{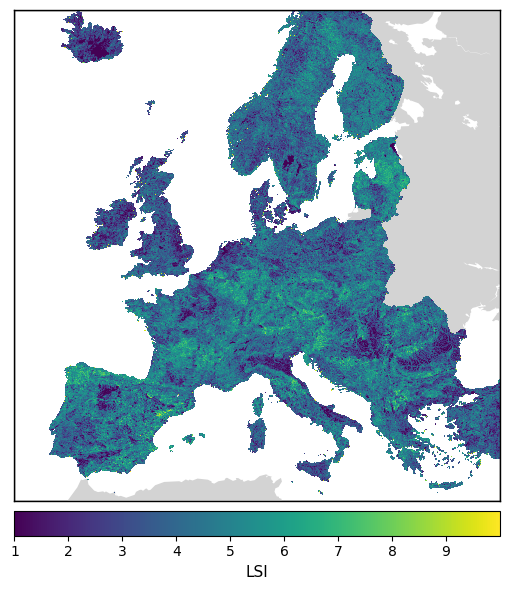} 
    \end{minipage}%
    \begin{minipage}[c]{0.32\textwidth}
        \centering
        MESH \
        \includegraphics[width=\textwidth, trim=0 20 0 0, clip]{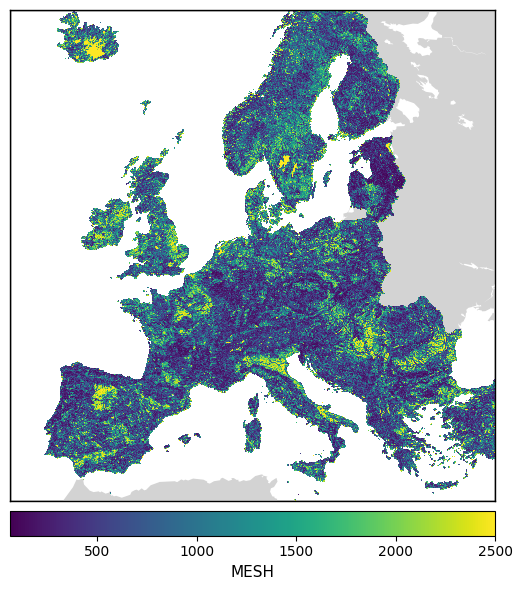} 
    \end{minipage}%
    \begin{minipage}[c]{0.32\textwidth}
        \centering
        LPI \
        \includegraphics[width=\textwidth, trim=0 20 0 0, clip]{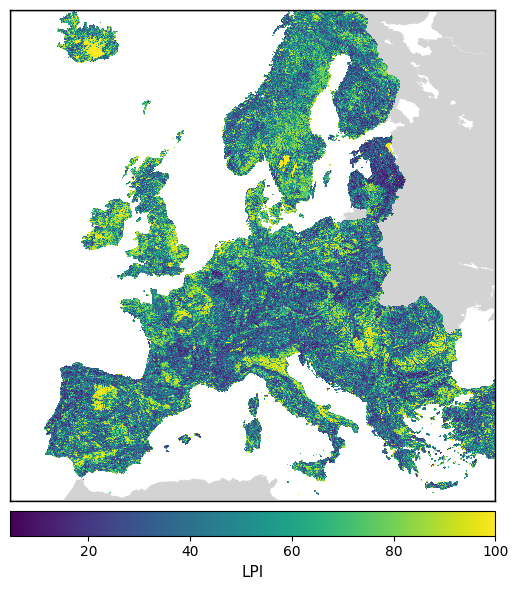} 
    \end{minipage}%
    
    \caption{Landscape metrics. From left to right: Landscape shape index (LSI), Effective mesh size (MESH), and Largest patch index (LPI).} 
    \label{fig:LandmMaps}
\end{figure}

Forest species diversity exhibits a pronounced spatial pattern along the latitudinal gradient, with lower diversity observed in southern Europe, maximal values in central regions, and a decline in northern latitudes (Figure~\ref{fig:Divmap}). Both the high severity areas of Southern -excluding Italy- and Northern Europe are characterized by lower species diversity. Landscape metrics maps are shown in Figure~\ref{fig:LandmMaps}, illustrating the variation of the spatial configuration patterns at the European scale. Heterogeneous land covers like shrubland, sparse vegetation, pastures and grasslands, and agroforestry generally exhibited high species diversity (Table~\ref{tab:Diversity_lcstats}). Among forest types, broad-leaved and mixed forests were found as the most diverse. Vineyards, agroforestry areas and mixed forests showed a higher shape and configurational complexity, reflecting greater diversity in patch shapes and connectivity. In contrast, shrubland, pastures, and coniferous forests generally showed lower spatial complexity (Tables~\ref{tab:LSI_lcstats}-\ref{tab:LPI_lcstats}).

\begin{figure}[H]
    \centering
    \begin{minipage}[c]{0.25\textwidth} 
        \centering
        Severity-Diversity \
        \includegraphics[width=\linewidth, trim=20 20 0 0, clip]{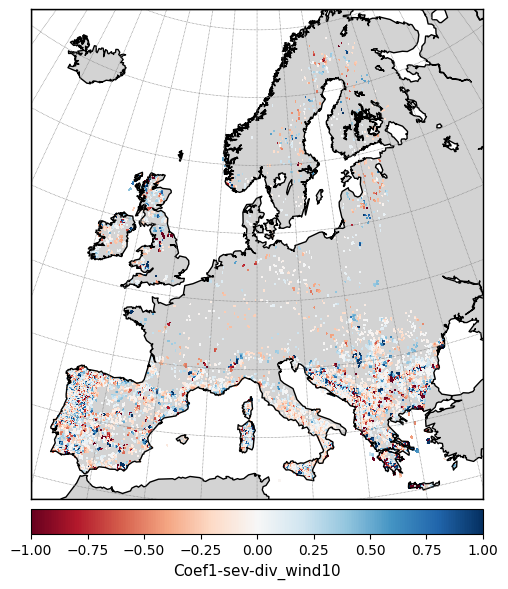} 
    \end{minipage}%
    \begin{minipage}[c]{0.25\textwidth}
        \centering
        Severity-LSI \
        \includegraphics[width=\linewidth, trim=0 20 0 0, clip]{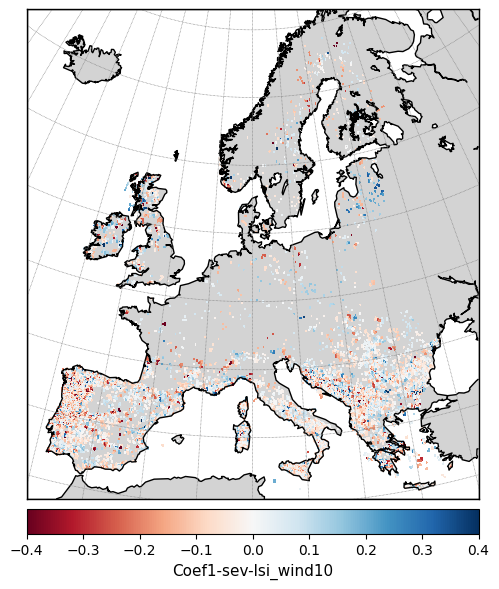} 
    \end{minipage}%
    \begin{minipage}[c]{0.25\textwidth}
        \centering
        Severity-MESH \
        \includegraphics[width=\linewidth, trim=0 20 0 0, clip]{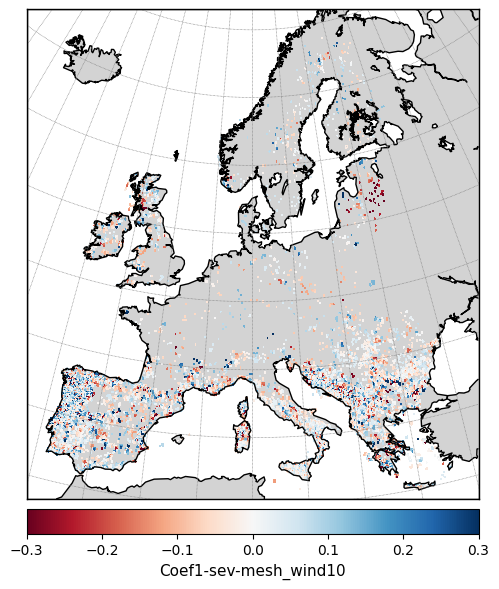} 
    \end{minipage}%
    \begin{minipage}[c]{0.25\textwidth}
        \centering
        Severity-LPI \
        \includegraphics[width=\linewidth, trim=0 20 0 0, clip]{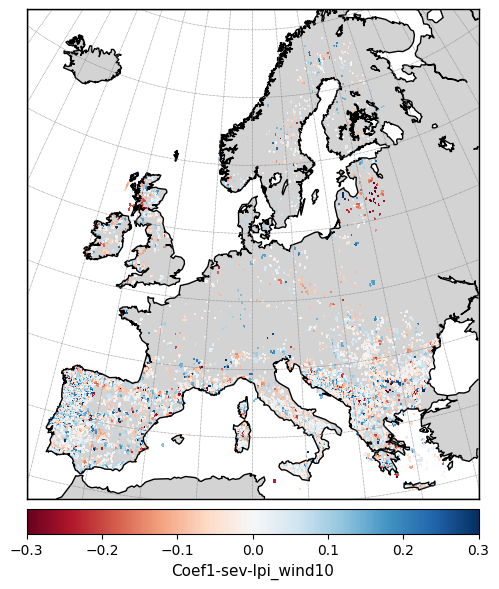} 
    \end{minipage}%
    
    \vspace{5mm}

    \begin{minipage}[c]{0.25\textwidth} 
        \centering
        Recovery-Diversity \
        \includegraphics[width=\linewidth,trim=0 20 0 0, clip]{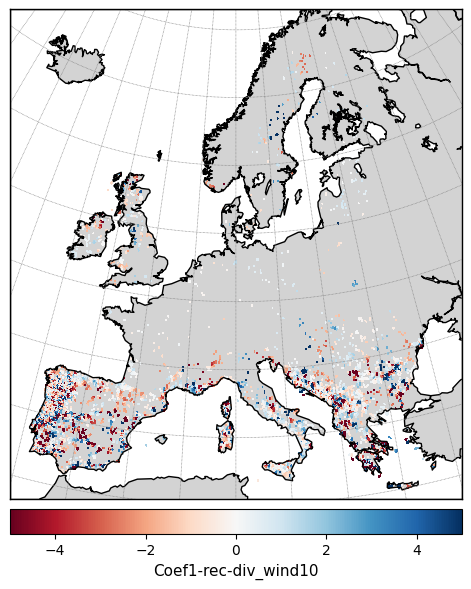} 
    \end{minipage}%
    \begin{minipage}[c]{0.25\textwidth}
        \centering
        Recovery-LSI \
        \includegraphics[width=\linewidth, trim=0 20 0 0, clip]{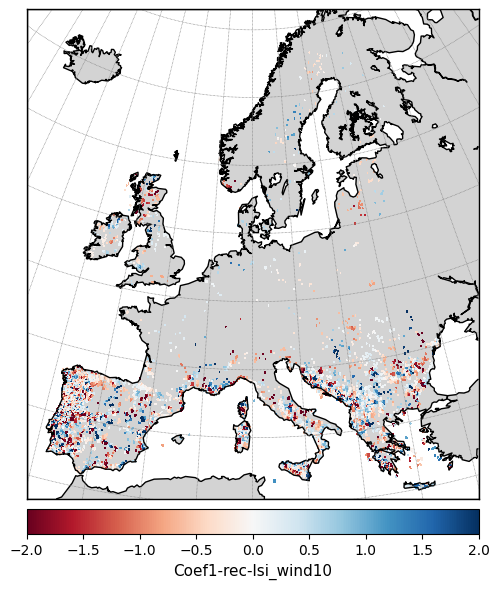} 
    \end{minipage}%
    \begin{minipage}[c]{0.25\textwidth}
        \centering
        Recovery-MESH \
        \includegraphics[width=\linewidth, trim=0 20 0 0, clip]{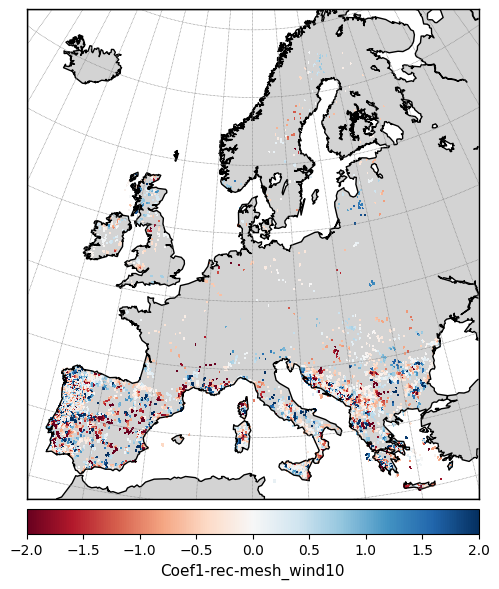} 
    \end{minipage}%
    \begin{minipage}[c]{0.25\textwidth}
        \centering
        Recovery-LPI \
        \includegraphics[width=\linewidth, trim=0 20 0 0, clip]{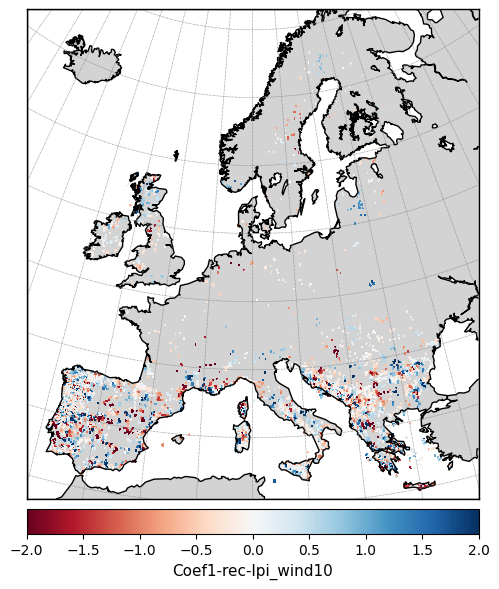} 
    \end{minipage}%

    \caption{Spatial distribution of geographically weighted regression (GWR) coefficients, illustrating the spatial variation in the relationship between fire severity (top row) and post-fire recovery time (bottom row) with forest species diversity and the landscape metrics.} 
    \label{fig:gwr-Coefs-sevrec}
\end{figure}

The spatial analysis using GWR was conducted to assess the relationship between fire severity and post-fire recovery with regard to forest species diversity and landscape heterogeneity across Europe. Separate GWR models were fitted independently for each fire metric --severity and recovery-- against species diversity and three landscape metrics (LSI, MESH, and LPI) (Table \ref{tab:GWRstats}). This approach enabled spatial comparison of regression coefficients, offering insights into localized correlations between variables (Figure \ref{fig:gwr-Coefs-sevrec}). Results show mainly negative coefficients between both fire metrics and forest diversity, particularly in southern European regions more frequently affected by fire. LSI consistently exhibited the strongest negative relationship, suggesting that more geometrically complex landscapes tend to experience lower severity and shorter recovery periods following fire disturbance. In contrast, MESH and LPI exhibit positive correlations with fire severity and recovery, suggesting that greater landscape connectivity and dominance are associated with higher fire impact. The observed correlations are generally stronger for recovery, as reflected overall in the higher coefficient values across Europe.

\begin{table}[H]
\begin{center}
\caption{Summary statistics of geographically weighted regression (GWR) and global regression estimate, with severity and recovery modeled as the dependent variable for each listed metric.} 
\resizebox{\textwidth}{!}{
\begin{tabular}{lccccccc}
\hline
\multicolumn{8}{c}{\textbf{Severity (-)}} \\
& $\text{Coeff}_{\text{GWR}}$ mean & $\text{Coeff}_{\text{GWR}}$ std & $\text{Coeff}_{\text{GWR}}$ median & $R^2_{\text{GWR}}$ & 
$\text{Coeff}_{\text{Global}}$ & $\text{t-statistic}_{\text{Global}}$ & $\text{p}_{\text{value}}$\\

\hline

Diversity & 0.001 & 1.467 & -0.007 & 0.439 & -0.031 & -47.521 & $<$0.001 \\

Landscape shape index (LSI) & -0.024 & 0.165 & -0.021 & 0.440 & -0.018 & -28.064 & $<$0.001\\

Effective mesh size (MESH) & 0.019 & 0.165 & 0.016 & 0.441 & 0.027 & 44.119 & $<$0.001\\

Largest patch index (LPI) & 0.020 & 0.121 & 0.015 & 0.446 & 0.028 & 44.398 & $<$0.001 \\

\hline

\multicolumn{8}{c}{\textbf{Recovery (years)}}\\
& $\text{Coeff}_{\text{GWR}}$ mean & $\text{Coeff}_{\text{GWR}}$ std & $\text{Coeff}_{\text{GWR}}$ median & $R^2_{\text{GWR}}$ & 
$\text{Coeff}_{\text{Global}}$ & $\text{t-statistic}_{\text{Global}}$ & $\text{p}_{\text{value}}$ \\
\hline

Diversity & -0.885 & 10.725 & -0.346 & 0.425 & -0.396 & -33.531 & $<$0.001\\
Landscape Shape Index (LSI) & -0.172 & 1.633 & -0.093  & 0.430 & -0.175 & -15.096 & $<$0.001\\
Effective mesh size (MESH) & 0.098 & 1.626	& 0.012	& 0.432	& 0.190 & 16.545 & $<$0.001\\
Largest Patch Index (LPI) & 0.107 & 1.243 & 0.034 & 0.435 & 0.180 & 15.591 & $<$0.001\\
\hline

\end{tabular}
}
\label{tab:GWRstats}
\end{center}
\end{table}

\begin{figure}[H]
    \centering
    \begin{minipage}[c]{0.5\textwidth} 
        \centering
        Severity \
        \includegraphics[width=\textwidth,trim=0 40 0 0, clip]{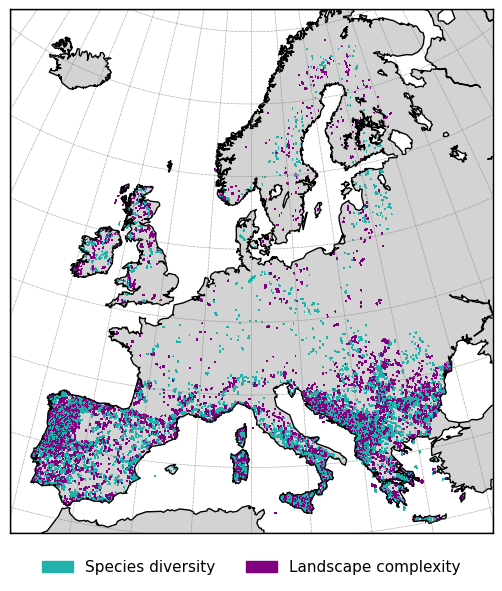} 
    \end{minipage}%
    \begin{minipage}[c]{0.5\textwidth}
        \centering
        Recovery \
        \includegraphics[width=\textwidth, trim=0 40 0 0, clip]{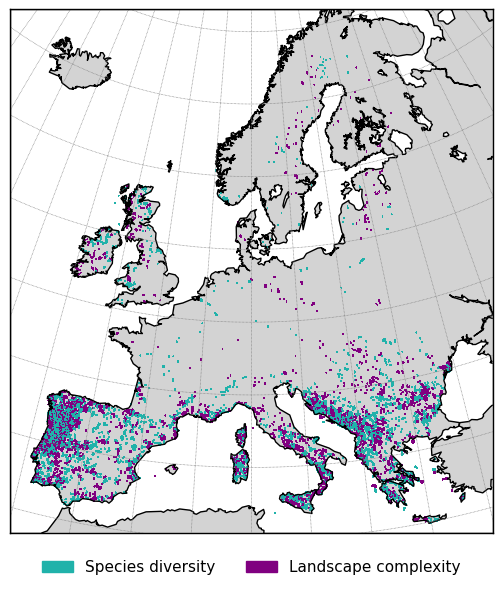} 
    \end{minipage}%
    
    \vspace{1em} 
    \centering
    \includegraphics[width=0.5\textwidth, trim=0 0 0 395, clip]{Coef12-rec-div-lsi_min_wind10_2var.png} 
    
    \caption{Mapping of the predominant factors influencing fire severity (left) and post-fire recovery time (right), as identified by the highest geographically weighted regression (GWR) coefficient per pixel. Each pixel is classified according to its dominant explanatory variable --either forest species diversity or landscape complexity-- highlighting regional variations in the severity and recovery processes.} 
    
    \label{fig:GWRsevrec-2var}
\end{figure}

To further identify the key drivers of fire severity and post-fire recovery dynamics from diversity and landscape shape complexity, a GWR model was fitted using both forest species diversity and LSI variables (Table \ref{tab:GWRstats2var}). LSI was selected based on their previously demonstrated correlation strength and to minimize redundancy in the information provided by the landscape metrics. The GWR coefficients for each variable were subsequently used to map their relative influence. The dominant factor was identified as the variable exhibiting the strongest negative correlation (Figure \ref{fig:GWRsevrec-2var}). The dominant variable driving both fire severity and recovery was assessed across land cover types by comparing the proportion of pixels where either forest species diversity or landscape complexity exhibited the strongest negative correlation (Table \ref{tab:SevRec_Div_LSI_2gwr_lcstats}). Results showed that, for severity, a greater proportion of pixels aligned with landscape complexity, while diversity emerged as the primary driver of recovery across all land cover classes.

\begin{table}[H]
\begin{center}
\caption{Summary statistics of geographically weighted regression (GWR) and global regression estimate, for fire severity and post-disturbance recovery, in relation to forest species diversity and landscape complexity.} 

\resizebox{\textwidth}{!}{
\begin{tabular}{lccccc}
\hline
\multirow{2}{*}{} & \multicolumn{5}{c}{\textbf{Severity (-)}} \\
& $\text{Coeff}_{\text{GWR}}$ median & $R^2_{\text{GWR}}$ & $\text{Coeff}_{\text{Global}}$ & $\text{t-statistic}_{\text{Global}}$ & $\text{p}_{\text{value}}$\\
\hline
Diversity & -0.011 & \multirow{2}{*}{0.487} & -0.031 & -47.728 & \multirow{2}{*}{$<$0.001}\\
Landscape shape index (LSI) & -0.020 & & -0.017 & -26.675 &  \\
\hline
\multirow{2}{*}{} & \multicolumn{5}{c}{\textbf{Recovery (years)}} \\
& $\text{Coeff}_{\text{GWR}}$ median & $R^2_{\text{GWR}}$ & $\text{Coeff}_{\text{Global}}$ & $\text{t-statistic}_{\text{Global}}$ & $\text{p}_{\text{value}}$\\
\hline
Diversity & -0.397 & \multirow{2}{*}{0.473} & -0.364 & -33.695  & \multirow{2}{*}{$<$0.001}\\
Landscape shape index (LSI) & -0.160 & & -0.109  & -13.604 &  \\
\hline

\end{tabular}
}
\label{tab:GWRstats2var}
\end{center}
\end{table}

\section{Discussion}
\doublespacing

Fire severity and subsequent recovery across Europe were characterized through mapping analyses conducted on fire events occurring from 2001 to 2024. Remote sensing-based recovery indicators are typically spectral-based, meaning they primarily reflect greening and overall vegetation presence rather than actual tree composition or biomass \citep{perez2021,chu2013}. The approach proposed in this study overcomes challenges in interpreting spectral vegetation indices by using LAI instead, allowing for an understanding of both measurable structural and greenness aspects of forest recovery. Compared to the commonly applied relative severity ratio, recovery estimates can vary significantly depending on the spectral indices and formulation used \citep{hislop2018,nole2022}. Compared to fixed threshold-based indicators, baseline recovery approaches, as adopted in this study, are calibrated by considering the pre-disturbance status of the forest, thus accounting for the typical long-term past canopy cover over different landscape contexts. To further ensure sustained forest recovery and avoid mistaking temporary post-fire greening for true regrowth, as well as considering the potential impact of subsequent threats, this study established a one-year recovery condition based on the baseline threshold.

Overall, a discernible pattern of severe fire intensity was observed in Southern Europe. High susceptibility to both an increased number of fire events and severe fire impacts reflects the confluence of climatic conditions typical of the Mediterranean climate and the prevalence of highly flammable vegetation types \citep{pausas2008, ermitao2021,li2025}. Shrublands, which experience the most frequent fires (Figure~\ref{fig:NfiresperLandCover}), predominantly exhibited high fire severity levels alongside coniferous and mixed forests. Both coniferous forests and shrubland also exhibited the longest recovery periods, consistent with prior studies \citep{white2022,ermitao2024}. This prolonged recovery can be attributed to the dominant pine woodland species found in these ecosystems, which primarily rely on seed banks for post-fire regeneration and are highly susceptible to fire in Mediterranean biomes (Figure~\ref{fig:fpsecies}) \citep{catry2010,maia2012,botequim2017}. Recurrent fires further exacerbate this vulnerability by promoting fire-prone vegetation, reinforcing fire regimes and hindering long-term recovery \citep{roche2024,thomsen2025}. Conversely, and in agreement with past research \citep{white2022,ermitao2024}, broad-leaved forests generally exhibited faster post-fire recovery. Prominent species in these forests include oak varieties common to temperate European regions (Figure~\ref{fig:fpsecies}), possessing traits favorable to fire adaptation such as resprouting and a protective bark \citep{annighofer2015,petersson2020}; and beech, though more susceptible to fire due to its typically thin bark, has demonstrated a relatively high regeneration capacity after fire \citep{maringer2020,moris2023}. Lower fire severity observed in vineyards, orchards, and agroforestry areas further supports the potential protective benefits of these land-use systems \citep{damianidis2021}, further enhanced by grazing livestock in silvopastoral systems that reduce fuel loads.

The GWR analysis further provided insights into the spatial patterns supporting the hypothesized inverse relationship between low species diversity, high fire severity, and extended recovery times. Species diversity showed a negative correlation with fire severity, aligning with previous studies that have demonstrated that forests with diverse species compositions are associated with reduced fire severity \citep{hely2000, wang2002, silva2009}. Also, ecosystem structural heterogeneity, including stand arrangement and functional diversity, has important implications for mitigating fire-induced mortality \citep{koontz2020}. In this study, landscape shape complexity --quantified by the Landscape shape index (LSI)-- was inversely related to fire severity, suggesting that fragmented, irregular landscapes help disrupt fuel continuity and reduce fire intensity. Conversely, positive correlations for MESH and LPI indicate that large, connected patches typical of homogeneous landscapes may exacerbate fire severity due to continuous fuel structures. GWR results indicated that landscape complexity had stronger local effects than global averages, revealing significant spatial variation in factors influencing fire severity. These patterns varied across regions, demonstrating the capacity of GWR to capture spatially variable contributions. Species diversity was the primary factor influencing fire severity across coniferous and mixed forests, vineyards, and agroforestry systems (Figure~\ref{fig:GWRsevrec-2var}, Table~\ref{tab:SevRec_Div_LSI_2gwr_lcstats}). These results suggest that the presence and composition of coniferous species directly influence fire severity in both pure and mixed stands. While forested areas tend to homogenize the landscape and increase fuel loads, severity in pastures and shrubland was more strongly shaped by landscape shape complexity. These areas, among the most susceptible to fires \citep{ganteaume2013}, typically border rural and natural vegetation sites, creating open canopies and a diverse mosaic of vegetation types that disrupt fuel continuity and hinder fire spread. Additionally, greater landscape complexity surrounding broad-leaved forests appears to contribute to reduced fire severity.

In addition to influencing forest susceptibility and resistance to disturbances, species composition can play a key role in post-disturbance recovery \citep{bartels2016, tepley2018}. Furthermore, landscape structural heterogeneity can substantially contribute to shaping post-disturbance recovery processes \citep{zaiats2024}. In this study, species diversity was the most influential factor driving forest recovery across Europe, outperforming landscape shape complexity. Higher diversity was consistently associated with faster vegetation re-establishment, particularly in broad-leaved and coniferous forests, likely due to reduced competition and favorable regeneration conditions. These effects were also evident in pastures, grasslands, and vineyards, where species diversity remained the predominant recovery factor. While landscape shape complexity supported recovery, especially in spatially fragmented areas, its influence was secondary to species composition. The inverse relationship between LSI and recovery time suggests that irregular landscape configurations facilitate regeneration by enhancing microhabitat diversity. Although the spatial variability of the GWR results, local configuration of more connectivity patterns less conducive to recovery. Despite the spatial variability in the GWR results, areas with more connected local configurations appeared generally less conducive to recovery.

Finally, this study assessed recovery solely through spectral greenness, overlooking forest attributes such as tree height and biomass, potentially underestimating recovery time \citep{bartels2016,viana2022}. This study also omitted site-specific factors, such as topography, climate, understory composition, and human pressures, which interact with forest diversity and landscape heterogeneity to shape fire impacts \citep{nunes2016,garcia2019,pausas2021}. Moreover, forest regrowth speed and success depend on disturbance severity, recurrence, and favorable site and climate conditions, with recovery trajectories varying across bioregions \citep{gordon2017,cerioni2024,nole2022}. Ultimately, the findings of this study underscore the ecological importance of preserving species-rich and structurally diverse landscapes to enhance post-fire resilience and facilitate sustained forest regeneration. In addition to these benefits, such landscapes can promote ecosystem multifunctionality and biodiversity conservation. Furthermore, maintaining mosaics of land cover types, along with the promotion of active land use that prevents abandonment, can contribute to the design of resilient landscapes to support ecological stability and ensure long-term sustainability. Future research should further assess ecosystem recovery in terms of biodiversity and key indicators of ecosystem functionality that underpin long-term resilience.

\section{Conclusions}
\doublespacing

This study implemented a remote sensing–based estimation approach that utilizes a moving average baseline threshold to estimate post-fire vegetation recovery, leveraging preceding vegetation conditions. Fires were more frequent in the southern Mediterranean region of Europe. Shrublands and coniferous forests exhibited the highest severity levels, with shrublands being the most affected and slowest to recover, followed by coniferous forests. In contrast, broad-leaved forests and agroforestry areas showed faster post-fire recovery. Across all vegetation cover groups, the observed trend suggests that low forest species diversity is associated with higher fire severity and, more significantly, with longer recovery periods. Similarly, landscape metrics revealed an overall negative association with severity and an even stronger connection to recovery, indicating that spatially complex landscape configurations may positively influence post-disturbance regeneration. Based on spatial patterns observed across Europe, landscape complexity appeared to play a more dominant role in fire severity, while species diversity had a greater influence on recovery outcomes.

\section{Acknowledgements}

This research was supported by the Horizon Europe project ECO2ADAPT, grant agreement no. 101059498.

\clearpage 
\bibliographystyle{elsarticle-harv} 
\bibliography{biblio.bib}

\clearpage 


\appendix
\renewcommand{\thesection}{S\arabic{section}}
\renewcommand{\thetable}{S\arabic{table}}
\renewcommand{\thefigure}{S\arabic{figure}}
\setcounter{section}{0}
\setcounter{table}{0}
\setcounter{figure}{0}

\section{Supplementary Material}

\subsection{Fires over Europe from 2000 to 2024}

\begin{figure}[H]
    \centering
    \includegraphics[width=\textwidth]{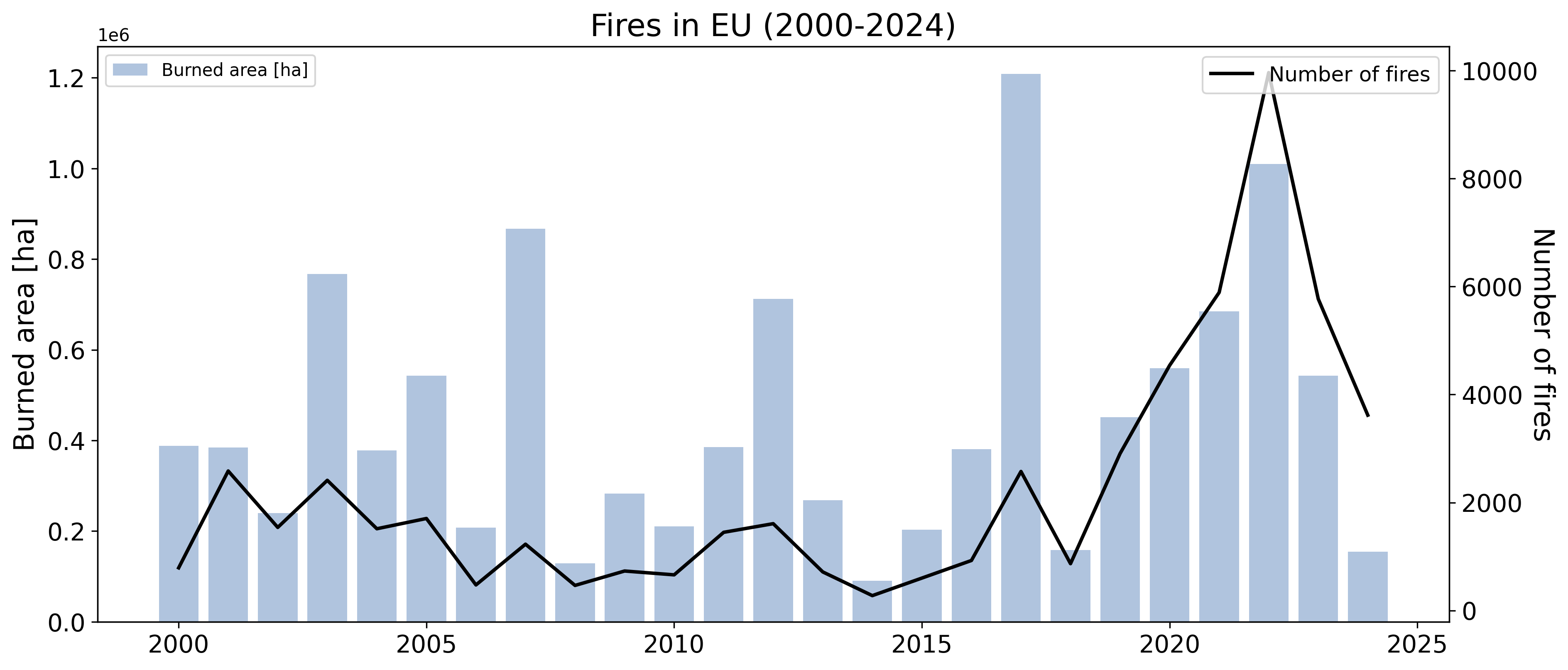}
        \caption{Number of fires and burned area over Europe from 2000 to 2024.}
    \label{fig:Nfires}
\end{figure}

\subsection{Fires over Europe from 2001 to 2024 per land cover class}

\begin{figure}[H]
    \centering
    \includegraphics[width=\textwidth,trim=15 0 0 0, clip]{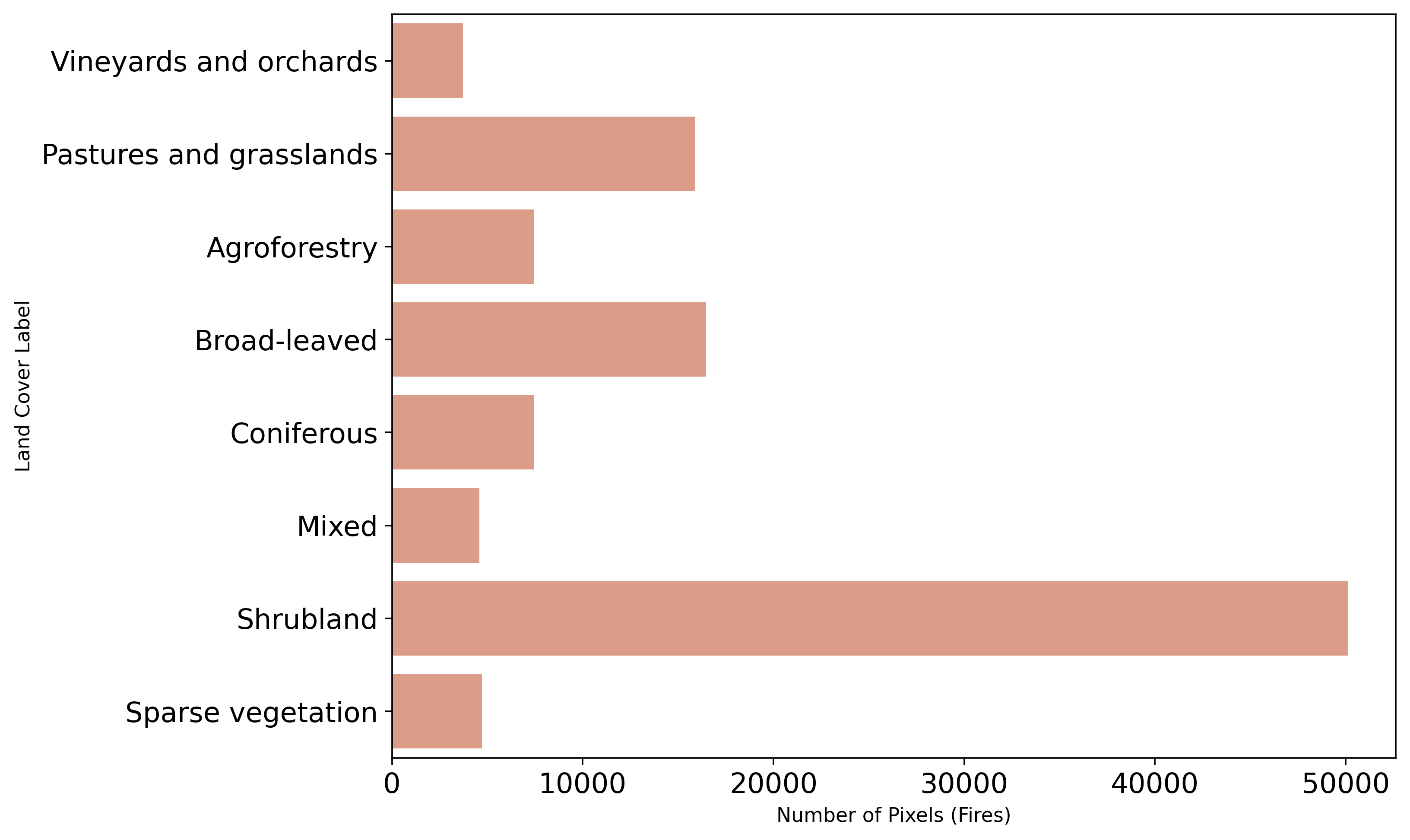}
        \caption{Fire pixel count over Europe from 2001 to 2024 per land cover class.}
    \label{fig:NfiresperLandCover}
\end{figure}

\subsection{CORINE reclassification}

\begin{table}[H]
\begin{center}
\caption{CORINE Land cover reclassification.} 
\scriptsize
\resizebox{\textwidth}{!}{
\begin{tabularx}{\textwidth}{|l|X|}
\hline
Reclassification land cover class & CORINE land cover class Level 3 \\ 
\hline
Vineyards and orchards & Vineyards \newline Fruit trees and berry plantations \newline Olive groves \\
\hline
Agroforestry  & Land principally occupied by agriculture, with significant areas of natural vegetation \newline Agroforestry areas \\ 
\hline
Pastures and grasslands & Pastures, meadows and other permanent grasslands under agricultural use \newline Natural grasslands \\
\hline
Shrubland & Moors and heathland \newline Sclerophyllous vegetation \newline Transitional woodland-shrub \\
\hline
Coniferous & Coniferous forest \\
\hline
Broad-leaved &  Broad-leaved forest\\
\hline
Mixed &  Mixed forest\\
\hline
Sparse vegetation &  Sparsely vegetated areas \\
\hline
Cropland &  Non-irrigated arable land  \newline Permanently irrigated land \newline Rice fields \newline Annual crops associated with permanent crops \newline Complex cultivation patterns \\
\hline
Urban &  Continuous urban fabric \newline Discontinuous urban fabric \newline Industrial or commercial units and public facilities \newline Road and rail networks and associated land \newline Port areas \newline Airports \newline Mineral extraction sites \newline Dump sites \newline Construction sites \newline Green urban areas \newline Sport and leisure facilities\\
\hline
No vegetated areas &  Beaches, dunes, sands \newline Bare rocks \newline Glaciers and perpetual snow \\
\hline
Wetlands &  Inland marshes \newline Peat bogs \newline Coastal salt marshes \newline Salines \\
\hline
Water bodies &  Intertidal flats \newline Water courses \newline Water bodies \newline Coastal lagoons \newline Estuaries \newline Sea and ocean \\
\hline
\end{tabularx}
}
\label{tab:LCReclass}
\end{center}
\end{table}

\subsection{Main forest species per land cover class}

\begin{figure}[H]
    \centering
    \begin{tabular}{cc}
        \includegraphics[width=0.5\textwidth]{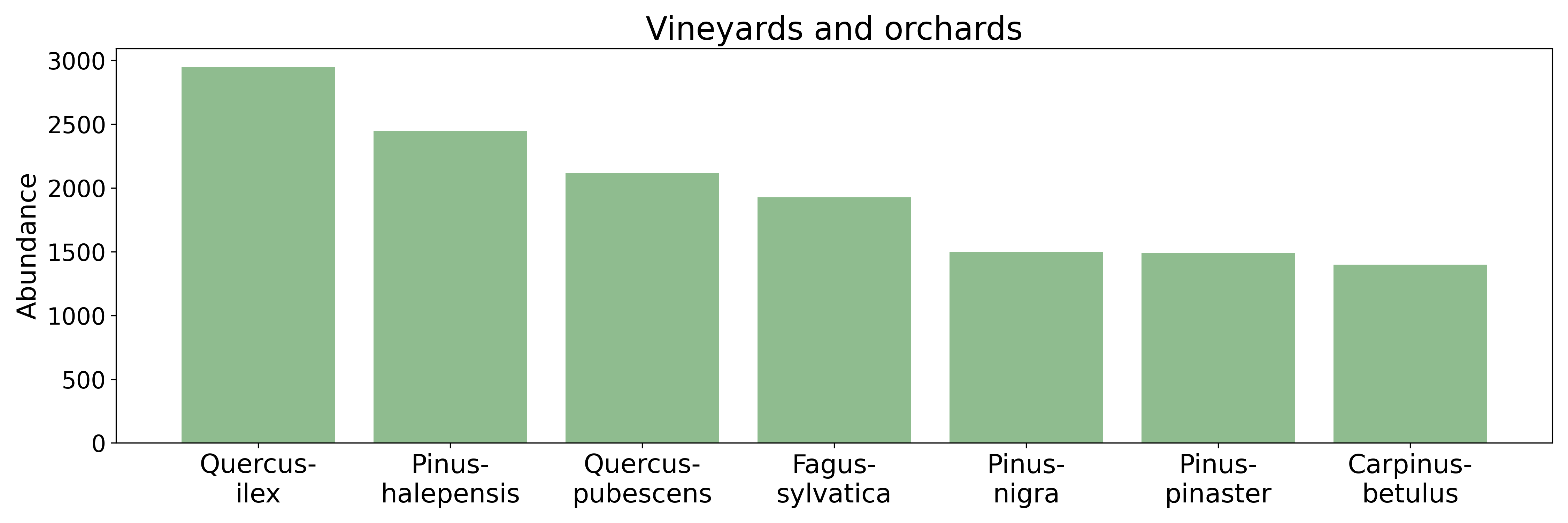} & \includegraphics[width=0.5\textwidth]{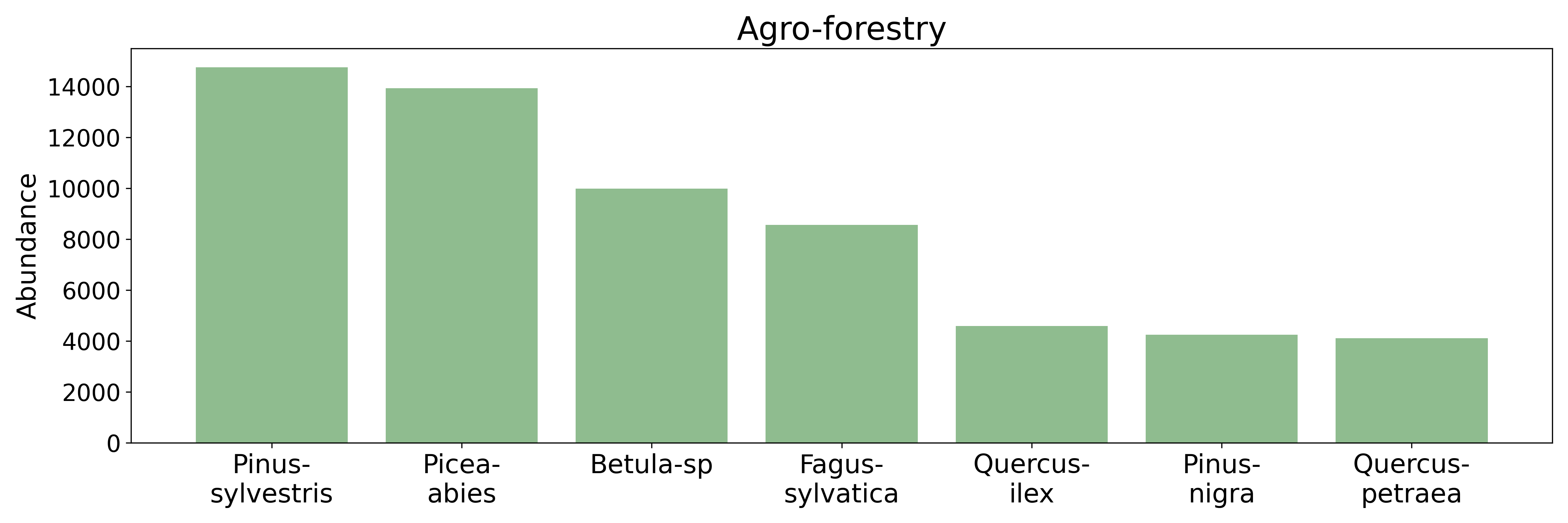} \\ 
        \includegraphics[width=0.5\textwidth]{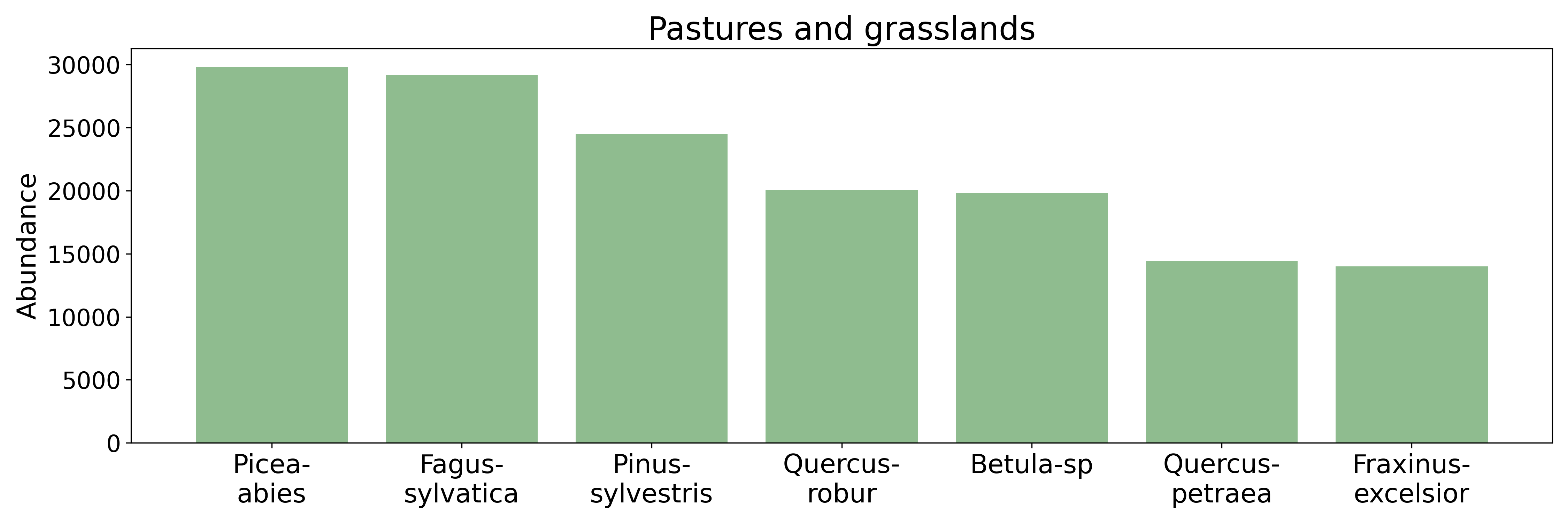} & \includegraphics[width=0.5\textwidth]{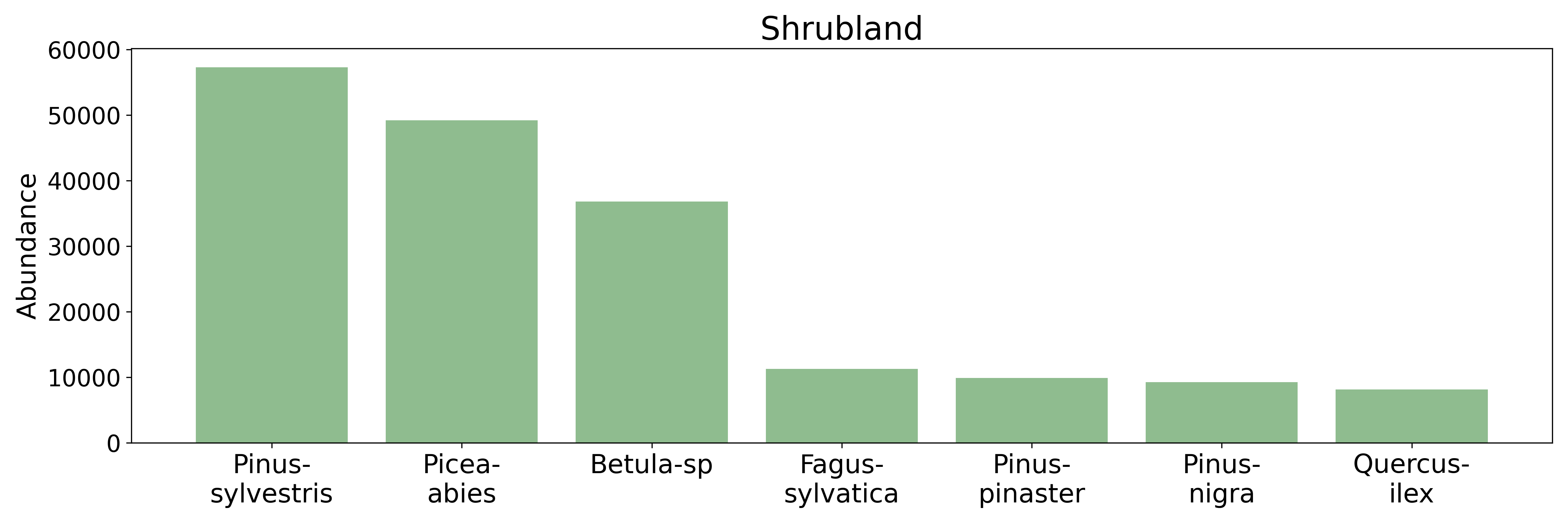} \\ 
        \includegraphics[width=0.5\textwidth]{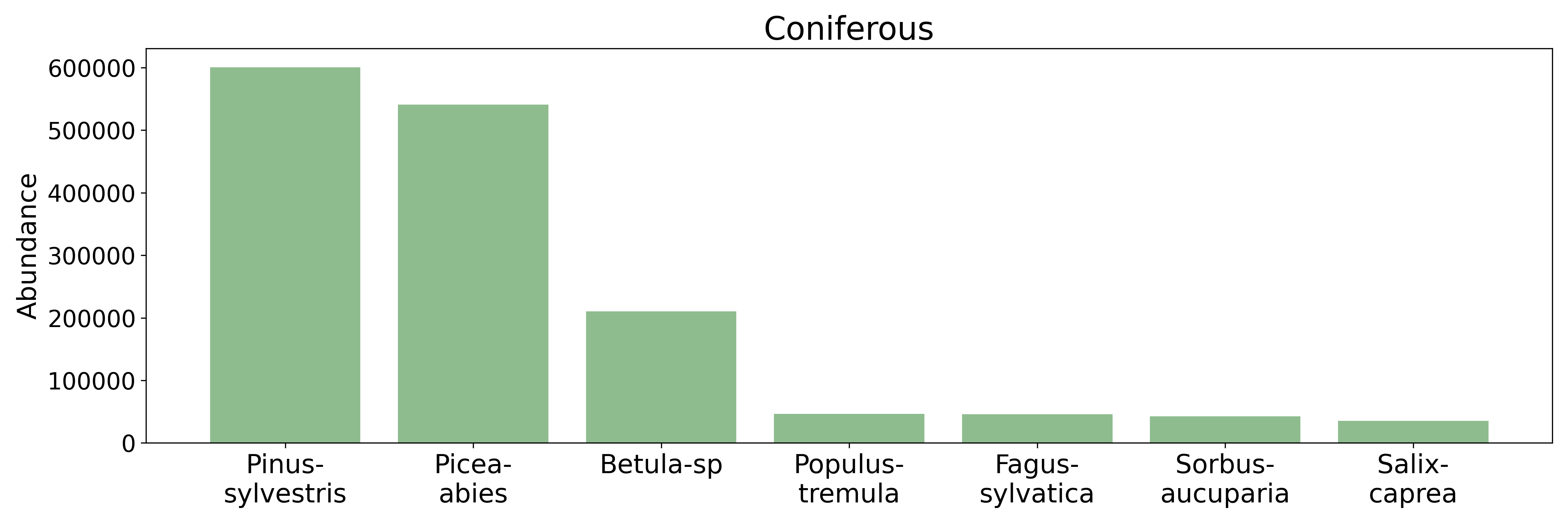} & \includegraphics[width=0.5\textwidth]{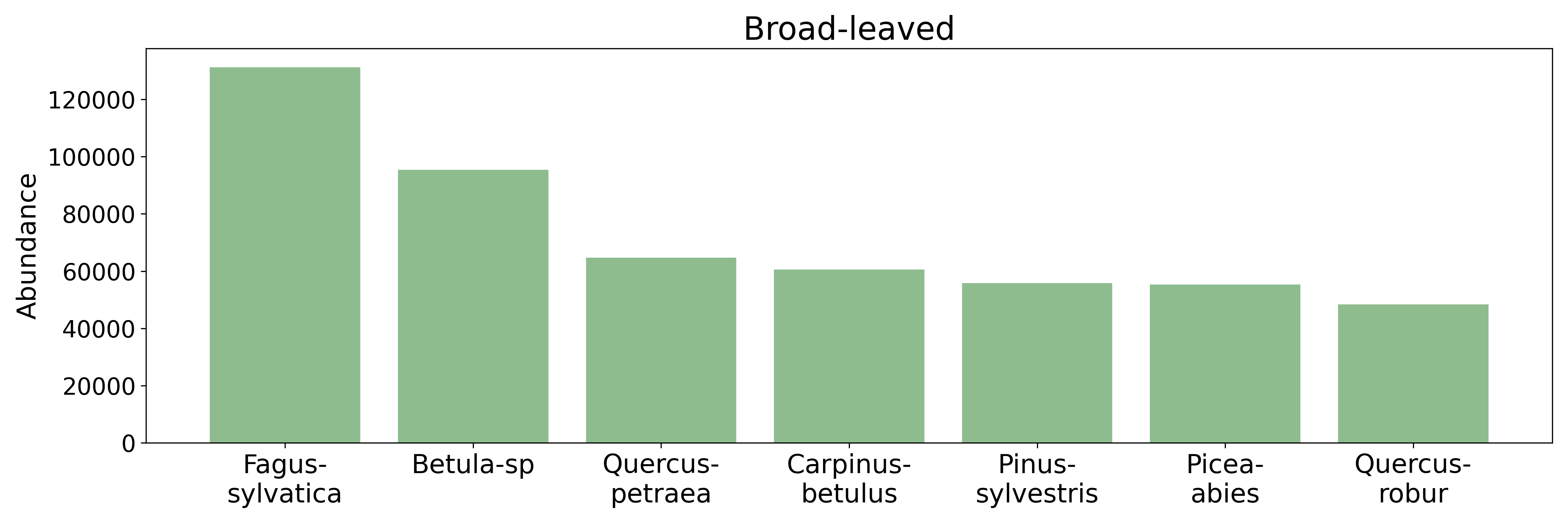} \\
        \includegraphics[width=0.5\textwidth]{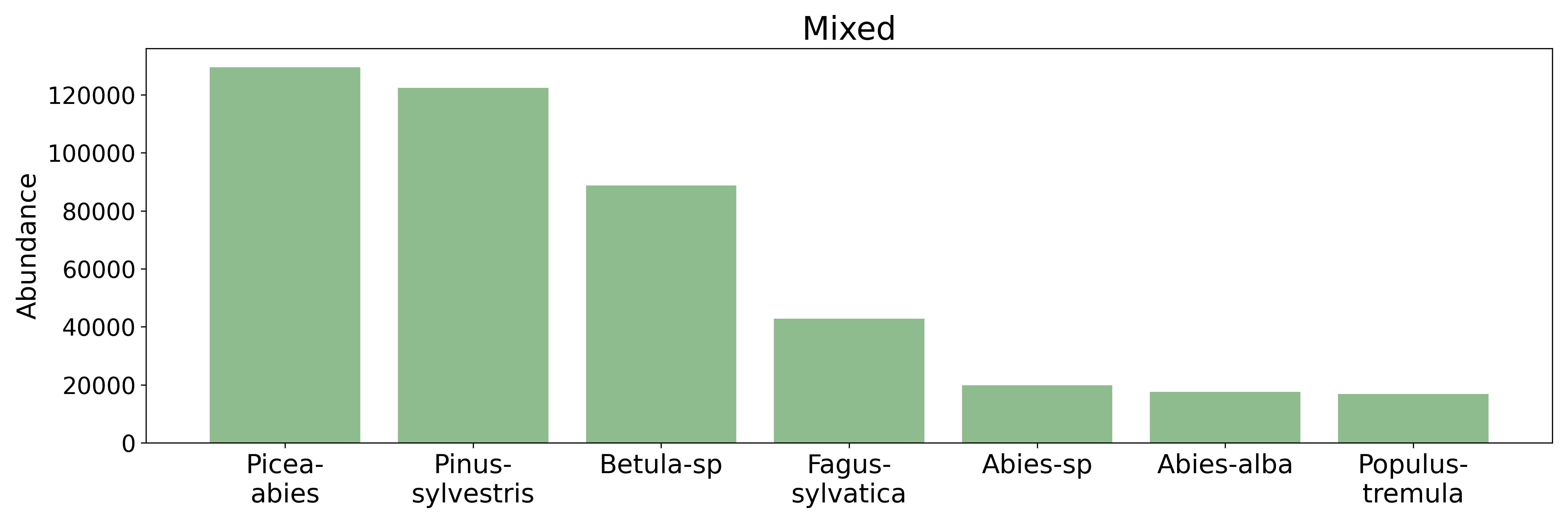} & \includegraphics[width=0.5\textwidth]{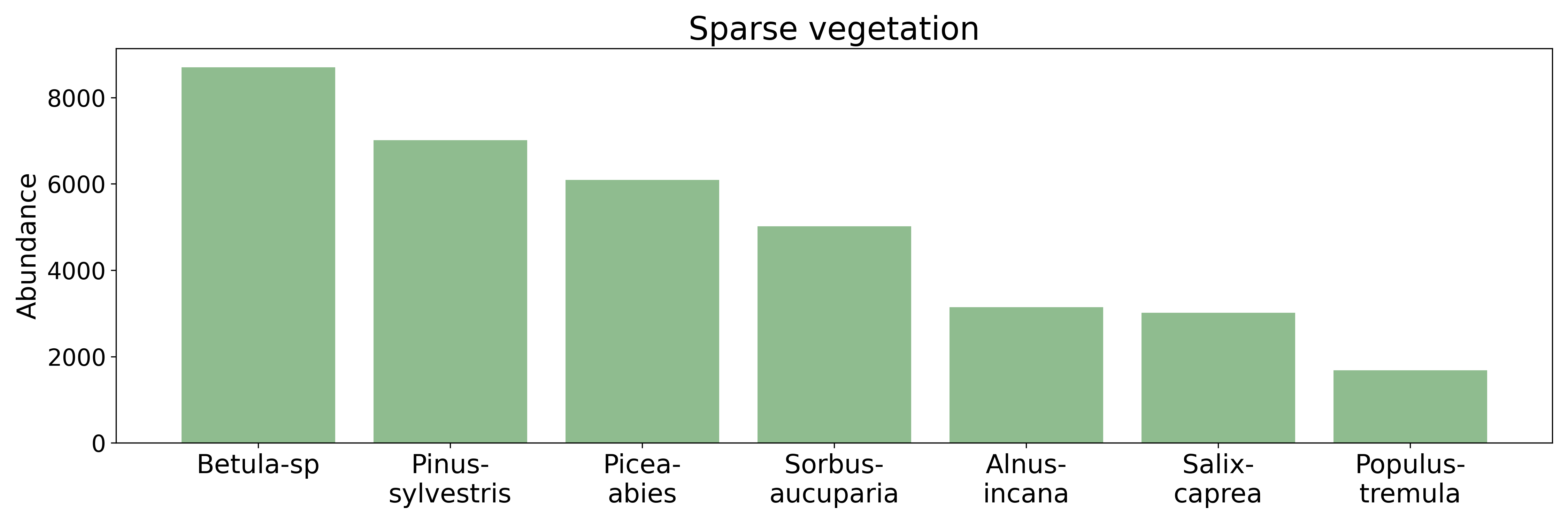}
    \end{tabular}
    \caption{Forest species abundance distribution per land cover class.}
    \label{fig:fpsecies}
\end{figure}

\subsection{Landscape metrics formulation}\label{sec:lmetrics_formulas}
\subsubsection{Landscape shape index (LSI)}
\begin{equation}
\begin{array}{c}
LSI = \dfrac{0.25E}{\sqrt{A}} \\
{\scriptstyle E: \text{ Total length of edge in landscape,}} \\
{\scriptstyle A: \text{ Total area}}
\end{array}
\end{equation}

\subsubsection{Effective mesh size (MESH)}
\begin{equation}
\begin{array}{c}
MESH = \frac{1}{A} \sum_{i=1}^{m} \sum_{j=1}^{n_i} a^{2}_{i,j}\\
{\scriptstyle A: \text{ Total area,}} \\
{\scriptstyle a_{i,j}: \text{ Area of patch ij}}
\end{array}
\end{equation}

\subsubsection{Largest patch index (LPI)}
\begin{equation}
\begin{array}{c}
LPI = \frac{1}{A} max \ a_{i,j}\\
{\scriptstyle A: \text{ Total area,}} \\
{\scriptstyle a_{i,j}: \text{ Area of patch ij}}
\end{array}
\end{equation}

\subsection{Forest species diversity and landscape metrics per land cover class}

\begin{table}[H]
\centering
\begin{tabularx}{\textwidth}{|X|c|c|c|c|}
\hline
\textbf{Land Cover Class} & \textbf{Mean} & \textbf{Std} & \textbf{Median} & \textbf{Mode} \\
\hline
    Vineyards and orchards    & 1.650 & 0.683 & 1.639 & 1.437 \\
    Agroforestry             & 1.657 & 0.562 & 1.673 & 1.891 \\
    Pastures and grasslands   & 1.913 & 0.542 & 1.938 & 2.148 \\
    Shrubland                 & 1.650 & 0.535 & 1.692 & 1.889 \\
    Coniferous                & 1.375 & 0.616 & 1.396 & 1.377 \\
    Broad-leaved              & 1.768 & 0.589 & 1.777 & 1.250 \\
    Mixed                     & 1.512 & 0.563 & 1.456 & 1.417 \\
    Sparse vegetation         & 1.814 & 0.576 & 1.907 & 1.984 \\
\hline
\end{tabularx}
\caption{Forest species diversity summary statistics per land cover class.}
\label{tab:Diversity_lcstats}
\end{table}

\begin{table}[H]
\centering
\begin{tabularx}{\textwidth}{|X|c|c|c|c|}
\hline
\textbf{Land Cover Class} & \textbf{Mean} & \textbf{Std} & \textbf{Median} & \textbf{Mode} \\
\hline
    Vineyards and orchards    & 4.773 & 1.137 & 4.751 & 4.835 \\
    Agroforestry             & 4.661 & 1.015 & 4.643 & 4.606 \\
    Pastures and grasslands   & 4.262 & 1.191 & 4.259 & 4.259 \\
    Shrubland                 & 4.283 & 1.138 & 4.279 & 4.291 \\
    Coniferous                & 4.464 & 1.208 & 4.456 & 4.151 \\
    Broad-leaved              & 4.642 & 1.071 & 4.618 & 4.595 \\
    Mixed                     & 4.724 & 1.068 & 4.802 & 5.005 \\
    Sparse vegetation         & 4.188 & 0.995 & 4.167 & 3.792 \\
\hline
\end{tabularx}
\caption{Landscape shape index (LSI) summary statistics per land cover class.}
\label{tab:LSI_lcstats}
\end{table}

\begin{table}[H]
\centering
\begin{tabularx}{\textwidth}{|X|c|c|c|c|}
\hline
\textbf{Land Cover Class} & \textbf{Mean} & \textbf{Std} & \textbf{Median} & \textbf{Mode} \\
\hline
    Vineyards and orchards    & 504.533 & 312.264 & 424.747 & 324.865 \\
    Agroforestry             & 481.849 & 292.201 & 406.266 & 286.451 \\
    Pastures and grasslands   & 683.785 & 435.348 & 561.118 & 367.734 \\
    Shrubland                 & 759.821 & 466.188 & 635.164 & 395.616 \\
    Coniferous                & 673.016 & 425.941 & 551.580 & 343.748 \\
    Broad-leaved              & 626.485 & 377.318 & 523.877 & 346.053 \\
    Mixed                     & 474.644 & 332.026 & 379.784 & 262.788 \\
    Sparse vegetation         & 595.629 & 354.638 & 507.838 & 387.858 \\
\hline
\end{tabularx}
\caption{Effective mesh size (MESH) summary statistics per land cover class.}
\label{tab:MESH_lcstats}
\end{table}

\begin{table}[H]
\centering
\begin{tabularx}{\textwidth}{|X|c|c|c|c|}
\hline
\textbf{Land Cover Class} & \textbf{Mean} & \textbf{Std} & \textbf{Median} & \textbf{Mode} \\
\hline
    Vineyards and orchards    & 34.447 & 14.670 & 31.971 & 28.142 \\
    Agroforestry             & 33.485 & 14.514 & 30.611 & 24.948 \\
    Pastures and grasslands   & 42.402 & 18.545 & 38.937 & 28.461 \\
    Shrubland                 & 46.627 & 19.603 & 44.034 & 32.759 \\
    Coniferous                & 42.319 & 18.885 & 38.936 & 29.372 \\
    Broad-leaved              & 40.795 & 17.268 & 37.824 & 28.626 \\
    Mixed                     & 33.087 & 15.689 & 29.584 & 22.845 \\
    Sparse vegetation         & 38.043 & 15.566 & 35.211 & 30.338 \\
\hline
\end{tabularx}
\caption{Largest patch index (LPI) summary statistics per land cover class.}
\label{tab:LPI_lcstats}
\end{table}

\subsection{Geographically Weighted Regression (GWR) coefficient summary statistics per land cover class}

\begin{table}[H]
\centering
\begin{tabularx}{\textwidth}{|X|c|c|c|c|}
\hline
\textbf{Land Cover Class} & \textbf{Mean} & \textbf{Std} & \textbf{Median} & \textbf{Mode} \\
\hline
    Vineyards and orchards    & -0.047 & 3.176 & -0.081 & -0.041 \\
    Agroforestry             & -0.070 & 1.640 & -0.025 & -0.019 \\
    Pastures and grasslands   & 0.074 & 1.663 & 0.024 & -0.029 \\
    Shrubland                 & 0.014 & 1.356  & 0.003 & 0.016 \\
    Coniferous                & -0.033 & 0.520  & -0.039 & -0.039 \\
    Broad-leaved              & -0.039 & 0.867 & -0.013 & -0.024 \\
    Mixed                     & -0.056 & 0.714 & -0.021 & -0.012 \\
    Sparse vegetation         & 0.038 & 1.313 & -0.008 & 0.000 \\
\hline
\end{tabularx}
\caption{GWR Severity-diversity fitting summary statistics per land cover class.}
\label{tab:Sev_Diversity_gwr_lcstats}
\end{table}

\begin{table}[H]
\centering
\begin{tabularx}{\textwidth}{|X|c|c|c|c|}
\hline
\textbf{Land Cover Class} & \textbf{Mean} & \textbf{Std} & \textbf{Median} & \textbf{Mode} \\
\hline
    Vineyards and orchards    & -0.006 & 0.154 & -0.010 & -0.011 \\
    Agroforestry             & -0.018 & 0.172 & -0.016 & -0.010 \\
    Pastures and grasslands   & -0.022 & 0.160 & -0.017 & -0.008 \\
    Shrubland                 & -0.032 & 0.169 & -0.029 & -0.017 \\
    Coniferous                & -0.022 & 0.173 & -0.018 & -0.010 \\
    Broad-leaved              & -0.019 & 0.156 & -0.014 & -0.003 \\
    Mixed                     & -0.021 & 0.184 & -0.028 & -0.033 \\
    Sparse vegetation         & -0.030 & 0.171 & -0.023 & -0.017 \\
\hline
\end{tabularx}
\caption{GWR Severity-LSI fitting summary statistics per land cover class.}
\label{tab:Sev_LSI_gwr_lcstats}
\end{table}

\begin{table}[H]
\centering
\begin{tabularx}{\textwidth}{|X|c|c|c|c|}
\hline
\textbf{Land Cover Class} & \textbf{Mean} & \textbf{Std} & \textbf{Median} & \textbf{Mode} \\
\hline
    Vineyards and orchards    & -1.492 & 27.330 & -0.423 & -0.666 \\
    Agroforestry             & -1.726 & 9.527 & -0.411 & -0.183 \\
    Pastures and grasslands   & -1.320 &  7.755 & -0.305 & -0.064 \\
    Shrubland                 & -0.870 & 12.394  & -0.364 & -0.055 \\
    Coniferous                & -0.745 & 3.996  & -0.547 & -0.457 \\
    Broad-leaved              & -0.498 & 4.928 & -0.158 & -0.094 \\
    Mixed                     & 0.013 & 7.362 & -0.443 & -0.368 \\
    Sparse vegetation         & -0.629 & 5.181  & -0.209 & -0.101 \\
\hline
\end{tabularx}
\caption{GWR Recovery-diversity fitting summary statistics per land cover class.}
\label{tab:Rec_Diversity_gwr_lcstats}
\end{table}

\begin{table}[H]
\centering
\begin{tabularx}{\textwidth}{|X|c|c|c|c|}
\hline
\textbf{Land Cover Class} & \textbf{Mean} & \textbf{Std} & \textbf{Median} & \textbf{Mode} \\
\hline
    Vineyards and orchards    & -0.070 & 1.622 & -0.097 & -0.112 \\
    Agroforestry             & -0.093 & 1.656 & -0.078 & -0.091 \\
    Pastures and grasslands   &  0.002 & 1.549 &  0.062 &  0.061 \\
    Shrubland                 & -0.236 & 1.667 & -0.125 & -0.085 \\
    Coniferous                & -0.284 & 1.487 & -0.201 & -0.132 \\
    Broad-leaved              & -0.116 & 1.909 & -0.046 &  0.019  \\
    Mixed                     & -0.138 & 1.651 & -0.137 & -0.158 \\
    Sparse vegetation         & -0.114 & 1.503 & -0.112 & -0.103 \\
\hline
\end{tabularx}
\caption{GWR Recovery-LSI fitting summary statistics per land cover class.}
\label{tab:Rec_LSI_gwr_lcstats}
\end{table}

\subsection{Dominant negative correlations of diversity and landscape complexity by land cover class for fire severity and recovery}

\begin{table}[H]
\centering
\begin{tabular}{|>{\raggedright\arraybackslash}p{4.8cm}|>{\centering\arraybackslash}p{2cm}|>{\centering\arraybackslash}p{2cm}|>{\centering\arraybackslash}p{2cm}|>{\centering\arraybackslash}p{2cm}|}

\hline
& \multicolumn{2}{c|}{Severity (-)} & \multicolumn{2}{c|}{Recovery (years)} \\
\textbf{Land Cover Class} & \textbf{Diversity (\%)} & \textbf{LSI (\%)} & \textbf{Diversity (\%)} & \textbf{LSI (\%)} \\
\hline
Vineyards and orchards      & 58.19 & 41.81 & 57.80 & 42.20 \\
Agroforestry                & 50.76 & 49.24 & 56.16 & 43.84 \\
Pastures and grasslands     & 45.79 & 54.21 & 57.43 & 42.57 \\
Shrubland                   & 47.47 & 52.53 & 54.25 & 45.75 \\
Coniferous                  & 52.98 & 47.02 & 58.79 & 41.21 \\
Broad-leaved                & 49.07 & 50.93 & 50.61 & 49.39 \\
Mixed                       & 51.09 & 48.91 & 58.49 & 41.51 \\
Sparse vegetation           & 49.26 & 50.74 & 52.12 & 47.88 \\
\textbf{Total}              & 48.95 & 51.05 & 54.95 & 45.05 \\
\hline
\end{tabular}
\caption{Percentage of pixels per land cover class where forest species diversity or landscape shape index (LSI) exhibited the strongest negative correlation with fire severity and post-fire recovery. Results are reported separately for severity and recovery, indicating the dominant explanatory variable across vegetation types and for the total.}
\label{tab:SevRec_Div_LSI_2gwr_lcstats}
\end{table}




\end{document}